\documentclass[10pt]{article}
\begin{document}
\centerline {{\large\bf Exterior and evolutionary skew-symmetric differential}}  
\centerline  {{\large\bf forms and their role in mathematical physics}}
\centerline {L.~I. Petrova }
\centerline{{\it Moscow State University, Russia, ptr@cs.msu.su}}

\bigskip

At present the theory of skew-symmetric exterior  differential forms has been 
developed [1--6]. The closed exterior differential forms possess the invariant  
properties that are of great functional and 
utilitarian importance. The operators of the exterior differential form theory 
lie at the basis of the differential and integral operators of the field theory. 

However, the theory of exterior differential forms, being invariant one, does 
not answer the questions related to the evolutionary processes. 

In the work the readers are introduced to the skew-symmetric 
differential forms that possess evolutionary properties [7,8].  
They were called evolutionary ones. 

The radical  distinction between the evolutionary forms and the exterior 
ones consists in the fact that the exterior differential forms are defined on 
manifolds with {\it closed metric forms}, whereas the evolutionary differential 
forms are defined on manifolds with {\it unclosed metric forms}. 
This difference forms lead to that the
mathematical tools of exterior and evolutionary forms appear to be
directly opposed. At the basis of the mathematical apparatus of exterior
forms there lie identical relations, conjugated operators, nondegenerate
transforms, whereas the basis of the mathematical apparatus of evolutionary
forms makes up nonidentical relations, nonconjugated operators, 
degenerate transforms. Thus they complement each other and make up 
a complete mathematical apparatus. This mathematical apparatus allows 
description of discrete transitions, 
quantum steps, evolutionary processes,  generation of various structures. 
These are radically new possibilities of the mathematical physics. 
 
A significance of skew-symmetric  differential forms for mathematical physics, 
is related to the fact that they reflect properties of the conservation laws. 
The apparatus of skew-symmetric differential forms allowed to explain a 
mechanism of evolutionary processes in material media and to disclose the role 
of the conservation laws in these processes. 
These processes lead to origination of physical structures from which the 
physical fields and manifolds are formed. This elucidates 
the causality of physical phenomena. 

\bigskip
{\large\bf Introduction}

Development of the differential and integral calculus in 17 centure gave 
rise to a development of mathematical physics. Differential equations, 
boundary and initial problems enable one to describe many physical processes 
and phenomena. However, since the beginning of the 20th century a close 
connection of the differential and integral calculus 
and differential equations with physics has become weak. In mathematical 
physics it  
has arose the problem of invariant (independent of the choice of the 
coordinate system) description of physical phenomena. As the result, formalisms 
based on the tensor, group, variational methods, on the theories 
of the symmetries, transforms and so on with the basic requirement of 
invariance have been developed in physics. In particular, it is most evident in 
branches of physics that are related to field theories. 

It turns out that the closed exterior differential forms possess by all 
the same properties (invariant, group, tensor, structural, and other ones) 
that lie at the basis that of all invariant approaches in the mathematical 
physics [9-13]. Invariant 
properties of closed exterior differential forms explicitly or implicitly 
manifest themselves essentially in all formalisms of the field theory, such 
as the Hamilton formalism, tensor approaches, group methods, quantum mechanics 
equations, the Yang-Mills theory and others. The exterior differential forms 
integrate the algebraic and geometric approaches in physics. 
 
However, the invariant methods of the mathematical physics with all their 
significance nevertheless do not solve all the problems of methematical physics. 
They do not answer the question of related to the evolutionary processes, 
origination of various physical structures, formatting physical fields, a 
causality of physical phenomena, and so on. To answer 
these questions one again is forced to apply to the differential methods of 
investigations of the mathematical physics. And for these investigations the 
evolutionary differential forms appear to be necessary. They make it possible 
to keep track of the conjugacy of the equations, which describe physical 
processes that are necessary for investigating the evolutionary processes. 

The existence of the skew-symmetric differential forms (forms with exterior 
multiplication), which possess the evolutionary 
properties, has been established by the author while studying the problems of 
stability and generation of various physical structures. Investigation of these 
problems showed that there exists an unique mathematical apparatus of differential 
forms that makes it possible to elucidate the mechanism of evolutionary processes.

\section{Properties and specific features of the exterior differential
forms}
The exterior differential form of degree $p$ ($p$-form on the differentiable 
manifold) can be written as [5,6,11]
$$
\theta^p=\sum_{i_1\dots i_p}a_{i_1\dots i_p}dx^{i_1}\wedge
dx^{i_2}\wedge\dots \wedge dx^{i_p}\quad 0\leq p\leq n\eqno(1.1)
$$
Here $a_{i_1\dots i_p}$ are the functions of the variables $x^{i_1}$,
$x^{i_2}$, \dots, $x^{i_p}$, $n$ is the dimension of space,
$\wedge$ is the operator of exterior multiplication, $dx^i$,
$dx^{i}\wedge dx^{j}$, $dx^{i}\wedge dx^{j}\wedge dx^{k}$, \dots\
is the local basis which satisfies the condition of exterior
multiplication:
$$
\begin{array}{l}
dx^{i}\wedge dx^{i}=0\\
dx^{i}\wedge dx^{j}=-dx^{j}\wedge dx^{i}\quad i\ne j
\end{array}\eqno(1.2)
$$
[From here on the symbol $\sum$ will be omitted and it will be
implied that the summation is performed over double indices.  Besides, the
symbol of exterior multiplication will be also omitted for the
sake of presentation convenience].

The local domains of manifold are the basis of the exterior
form. 

It should be emphasized that the exterior differential forms and their
operators are considered locally.

The differential of the (exterior) form $\theta^p$ is expressed as
$$
d\theta^p=\sum_{i_1\dots i_p}da_{i_1\dots
i_p}dx^{i_1}dx^{i_2}\dots dx^{i_p} \eqno(1.3)
$$

From definition of a differential one can see that, firstly, the
differential of the exterior form is also the exterior form
(but with the degree $(p+1)$), and, secondly, he can see
that the components of the differential form commutator are
coefficients of the form differential. Thus, the differential
of the first-degree form $\omega=a_i dx^i$ can be written as
$d\omega=K_{ij}dx^i dx^j$ where $K_{ij}$ are the components of the
commutator for the form $\omega$ that are defined as
$K_{ij}=(\partial a_j/\partial x^i-\partial a_i/\partial x^j)$.

\subsection*{Closed exterior differential forms}
In physical applications the closed differential forms with
invariant properties appear to be of greatest practical utility.

A form is called a closed one if its differential is equal to zero:
$$
d\theta^p=0\eqno(1.4)
$$

From condition (1.4) one can see that the closed form is a conservative 
quantity. This means that it can correspond to the conservation law, namely, 
to some conservative physical quantity.

The differential of the form is a closed form. That is
$$
dd\omega=0\eqno(1.5)
$$
where $\omega$ is an arbitrary exterior form.

The form which is the differential of some other form 
$$
\theta^p=d\theta^{p-1}\eqno(1.6)
$$
is called an exact form. The exact forms prove to be closed
automatically [9]
$$
d\theta^p=dd\theta^{p-1}=0\eqno(1.7)
$$

Here it is necessary to pay attention to the following points. In the
above presented formulas it was implicitly assumed that the differential
operator $d$ is the total one (that is, the
operator $d$ acts everywhere in the vicinity of the point
considered locally),  and therefore it acts on the manifold of the
initial dimension $n$. However, the differential may be internal. Such a
differential acts on some structure with the dimension being less
than that of the initial manifold. The structure, on which the exterior
differential form may become a closed inexact form, is a pseudostructure
with respect to its metric properties. \{Cohomology, sections of cotangent 
bundles, the eikonal surfaces and so on may be regarded as examples of 
pseudostructures.\} The properties of pseudostructures will be considered
later.

If the form is closed on a pseudostructure only, the closure
condition is written as
$$
d_\pi\theta^p=0\eqno(1.8)
$$
And the pseudostructure $\pi$ obeys the condition
$$
d_\pi{}^*\theta^p=0\eqno(1.9)
$$
where ${}^*\theta^p$ is the dual form. 
(For the properties of dual forms see [12]).

From conditions (1.8) and (1.9) one can see that the form
closed on pseudostructure (a closed inexact form) is a conservative object, 
namely, this quantity conserves on pseudostructure. This can also correspond to
some conservation law, i.e. to conservative object.
(These are precisely the objects which are physical structures that 
form physical fields.)

The exact form is, by definition, a differential (see condition (1.6)).
In this case the differential is total. The closed inexact form is 
a differential too. The closed inexact form is an interior (on
pseudostructure) differential, that is
$$
\theta^p_\pi=d_\pi\theta^{p-1}\eqno(1.10)
$$

And so, any closed form is a differential of the form of a lower
degree: the total one $\theta^p=d\theta^{p-1}$ if the form is exact,
or the interior one $\theta^p=d_\pi\theta^{p-1}$ on pseudostructure if
the form is inexact. From this it follows that the form of lower
degree may correspond to the potential, and the closed form by itself
may correspond to the potential force. 

\subsection*{Invariant properties of closed exterior differential
forms. Conjugacy of the exterior differential forms}
Since the closed form is a differential (a total one if the form is exact,
or an interior one on the pseudostructure if the form is inexact), then it is
obvious that the closed form proves to be invariant under all
transforms that conserve the differential. The unitary transforms
(0-form), the tangent and canonical transforms (1-form),the gradient and gauge
transforms (2-form) and so on are  examples of such transforms.
{\it These are gauge transforms for spinor, scalar, vector, tensor (3-form)
fields}. (It is to be pointed out that just such transforms are used in field 
theory).

As mentioned above, from the closure condition it follows
that the closed exterior differential forms are conservative objects. 
As the result, the closed form is a conservative invariant object. The
covariance of the dual form is directly connected with the invariance
of the exterior closed inexact form. This property  of closed exterior 
differential forms and dual forms plays an essential role in describing the
conservation laws and lies at the basis of the field theory. 

The closure of the exterior differential forms and hence their
invariance result from the conjugacy of elements of the exterior or dual forms.

From the definition of the exterior differential form one can see that the
exterior differential forms have complex structure. The specific
features of the exterior form structure are homogeneity with
respect to the basis, skew-symmetry, the integration of terms
each consisting of two objects of different nature
(the algebraic nature for the form coefficients, and the geometric nature
for the base components). Besides,  the exterior form depends
on the space dimension and on the manifold topology. The closure
property of the exterior form means that any objects, namely,
elements of the exterior form, components of elements, elements of
the form differential, exterior and dual forms and others, turn
out to be conjugated. It is the conjugacy that leads to realization of
the invariant and covariant properties of the exterior and dual
forms that have a great functional and applied importance. The
variety of objects of conjugacy leads to the fact that the closed forms
can describe a great number of different physical and spatial
structures, and this fact once again emphasizes great mathematical
potentialities of the exterior differential forms.

\subsection*{Operators of the theory of exterior differential
forms}
In  differential calculus the derivatives are basic elements of
the mathematical apparatus. By contrast, the 
differential is an element of mathematical apparatus of the
theory of exterior differential forms. It enables one to analyze
the conjugacy of derivatives in various directions, which extends
potentialities of differential calculus. 

The operator of exterior differential $d$ (exterior
differential) is an abstract generalization of ordinary mathematical
operations of the gradient, curl, and divergence in the vector
calculus [6]. If, in addition to the exterior differential, we
introduce the following operators: (1) $\delta$ for transforms that convert
the form of $p+1$ degree into the form of $p$ degree, (2) $\delta'$
for cotangent transforms, (3) $\Delta$ for the
$d\delta-\delta d$ transform, (4)$\Delta'$ for the $d\delta'-\delta'd$
transform, then in terms of these operators that act on the exterior
differential forms one can write down the operators in the field
theory equations. The operator $\delta$ corresponds to Green's
operator, $\delta'$ does to the canonical transform operator,
$\Delta$ does to the d'Alembert operator in 4-dimensional space, and
$\Delta'$ corresponds to the Laplace operator [11-13]. It can be seen
that the operators of the exterior differential form theory are
connected with many operators of mathematical physics.

\subsection*{Identical relations of exterior differential forms}
In the theory of exterior differential forms the closed forms that
possess various types of conjugacy play a principal role. Since
the conjugacy is a certain connection between two operators or
mathematical objects, it is evident that relations can  be used to express
conjugacy mathematically. Just such relations constitute the basis of
the mathematical apparatus of the exterior differential forms. This is 
an identical relation.

The identical relations for exterior differential forms reflect the
closure conditions of the differential forms, namely, vanishing the form
differential (see formulas (1.4), (1.8), (1.9)) and the conditions
connecting the forms of consequent degrees (see formulas (1.6), (1.10)).

The importance of the identical relations for exterior differential forms
is manifested by the fact that practically in all branches of physics,
mechanics, thermodynamics one faces such identical relations. One can present 
the following examples:  

a) the Poincare invariant $ds\,=\,-H\,dt\,+\,p_j\,dq_j$,

b) the second principle of thermodynamics $dS\,=\,(dE+p\,dV)/T$,

c) the vital force theorem in theoretical mechanics: $dT=X_idx^i$
where $X_i$ are the components of the potential force, and $T=mV^2/2$ is the
vital force,

d) the conditions on characteristics [14] in the theory of differential
equations, and so on.
 
The identical relations in differential forms express the fact that each 
closed exterior form is a differential of some exterior form (with the degree 
less by one). In general form such an identical relation can be written as
$$
d _{\pi}\phi=\theta _{\pi}^p\eqno(1.11)
$$
In this relation the form in the right-hand side has to be a {\it closed } 
one. (As it will be shown below,
the identical relations are satisfied only on pseudostructures).

In identical relation (1.11) in one side it stands the closed form and
in other side does
the  differential of some differential form of the less by one degree, 
which is the closed form as well. 

In addition to relations in the differential forms from  
the closure conditions of the differential forms and the conditions
connecting the forms of consequent degrees  the identical relations of 
other types are obtained. The types of such relations are presented below. 

1. {\it Integral identical relations}.

The formulas by Newton, Leibnitz, Green, the integral relations by Stokes,
Gauss-Ostrogradskii are examples of integral identical relations. 

2. {\it Tensor identical relations}.

From the relations that connect exterior forms of consequent degrees
one can obtain the vector and tensor identical relations that connect
the operators of the gradient, curl, divergence and so on.

From the closure conditions of exterior and dual forms one can obtain the 
identical relations such as the gauge relations in electromagnetic field 
theory, the tensor relations between connectednesses and their derivatives
in gravitation (the symmetry of connectednesses with respect to lower indices,
the Bianchi identity, the conditions imposed on the Christoffel symbols) 
and so on.

3. {\it Identical relations between derivatives}.

The identical relations between derivatives correspond to the closure
conditions of exterior and dual forms. The examples of such relations are
the above presented Cauchi-Riemann conditions in the theory of complex
variables, the transversality condition in the calculus of variations,
the canonical relations
in the Hamilton formalism, the thermodynamic relations between derivatives
of thermodynamic functions [15], the condition that the derivative of
implicit function is subjected to, the eikonal relations [16] and so on.

\subsection*{Nondegenerate transforms}
One of the fundamental methods in the theory of exterior
differential forms is application of {\it nondegenerate}
transforms (below it will be shown that {\it degenerate} transforms
appear in the mathematical apparatus of the evolutionary forms).

Nondegenerate transforms, if applied to identical relations, enable one
to obtain new identical relations and new closed exterior differential forms.

In the theory of exterior differential forms the nondegenerate transforms 
are those that conserve the differential. 
The examples of nondegenerate transforms are unitary, tangent, canonical,
gradient, and gauge transforms.  

From description of operators of exterior differential forms one can see that
those are operators that execute some transforms. All these transforms
are connected with the above listed nondegenerate transforms of exterior
differential forms.

Thus, even from this brief description of the properties and
specific features of the exterior differential forms can clearly see wide 
functional and utilitarian potentialities of exterior differential forms. 

Before going to description of a new evolutionary differential forms it should 
dwell on the properties of manifolds on which skew-symmetric differential 
forms are defined.

\section{Some properties of manifolds}
In the definition of exterior differential forms a differentiable manifold 
was mentioned. Differentiable manifolds
are topological spaces that locally behave like Euclidean spaces [9,17].

But differentiable
manifolds are not a single type of manifolds on which the exterior
differential forms are defined. In the general case there are
manifolds with structures of any types. The theory of exterior
differential forms was developed just for such manifolds. They may
be the Hausdorff manifolds, fiber spaces, the comological,
characteristical, configuration manifolds and so on. These manifolds
and their properties are treated in [2,4,9,11] and in
some other works. Since all these manifolds possess
structures of any types, they have one common property, namely,
locally they admit one-to-one mapping into the Euclidean subspaces
and into other manifolds or submanifolds of the same dimension [9].

While describing the evolutionary processes in material system
one is forced to deal with manifolds which do not allow one-to-one
mapping described above. 
The manifolds that can be called accompanying manifolds are
variable manifolds. The differential forms that arise when evolutionary
processes are being described can be defined on manifolds of this type.

What are the characteristic
properties and specific features of accompanying manifolds and of
evolutionary differential forms connected with them?

To answer this question, let us analyze some properties of metric forms.

Assume that on the manifold one can set the
coordinate system with base vectors $\mathbf{e}_\mu$ and define
the metric forms of manifold [18]: $(\mathbf{e}_\mu\mathbf{e}_\nu)$,
$(\mathbf{e}_\mu dx^\mu)$, $(d\mathbf{e}_\mu)$. The metric forms
and their commutators define the metric and differential
characteristics of the manifold.

If metric forms are closed
(the commutators are equal to zero), then the metric is defined
$g_{\mu\nu}=(\mathbf{e}_\mu\mathbf{e}_\nu)$ and the results of
translation over manifold of the point
$d\mathbf{M}=(\mathbf{e}_\mu dx^\mu)$ and of the unit frame
$d\mathbf{A}=(d\mathbf{e}_\mu)$ prove to be independent of the
curve shape (the path of integration).

The closed metric forms
define the manifold structure, and the commutators of metric forms
define the manifold differential characteristics that specify
the manifold deformation: bending, torsion, rotation, twist.

It is evident that manifolds that are metric ones or possess the
structure have closed metric forms. It is with such manifolds that the
exterior differential forms are connected.

If the manifolds are deforming manifolds, this means that their
metric form commutators are nonzero. That is, the metric forms of such
manifolds turn out to be unclosed.
The accompanying manifolds appearing to be deforming are the
examples of such manifolds.

The skew-symmetric evolutionary differential forms 
whose basis are accompanying deforming manifolds are defined
on manifolds with unclosed metric forms.

What are the characteristic properties and specific features of such
manifolds and the related differential forms?

For description of the manifold differential characteristics
and, correspondingly, the metric forms commutators one can use
the connectednesses [2,5,9,18].

Let us consider the affine connectednesses and their relations to
commutators of metric forms.

The components of metric forms can be
expressed in terms of connectedness $\Gamma^\rho_{\mu\nu}$ [18]. The
expressions $\Gamma^\rho_{\mu\nu}$,
$(\Gamma^\rho_{\mu\nu}-\Gamma^\rho_{\nu\mu})$,
$R^\mu_{\nu\rho\sigma}$ are components of the commutators of
the metric forms of zero- first- and third degrees. (The commutator of
the second degree metric form is written down in a more complex
manner [18], and therefore it is not given here).

As it is known [18],
for the Euclidean manifold these commutators vanish identically.
For the Riemann manifold the commutator of the third-degree metric
form is nonzero: $R^\mu_{\nu\rho\sigma}\ne 0$. 

Commutators of interior metric forms vanish in the case of manifolds
that allow local one-to-one mapping into subspaces of
the Euclidean space. In other words, the metric forms of such
manifolds turn out to be closed.

With the metric form commutators the topological properties of manifolds
are connected. The metric form commutators
specify the manifold distortion. For example, the commutator of
the zero degree metric form $\Gamma^\rho_{\mu\nu}$ characterizes the
bend, that of the first degree form
$(\Gamma^\rho_{\mu\nu}-\Gamma^\rho_{\nu\mu})$ characterizes the torsion,
the commutator of the third -degree metric form $R^\mu_{\nu\rho\sigma}$
determines the curvature.

As it will be shown in the next section, it is with the properties of the
metric form commutators that the characteristic properties
of differential forms whose basis are manifolds with unclosed
metric forms are connected.

\section{Evolutionary differential forms}
Thus, the exterior differential forms are skew-symmetric differential forms  
defined on manifolds, submanifolds or on structures with closed
metric forms. The evolutionary differential forms are skew-symmetric 
differential forms defined on manifolds with metric forms that are unclosed. 

Let us point cut some properties of evolutionary forms and show what their
difference from  exterior differential forms is manifested in. 
An evolutionary differential form of degree $p$ ($p$-form),
as well as an exterior differential form, can be written down as
$$
\omega^p=\sum_{\alpha_1\dots\alpha_p}a_{\alpha_1\dots\alpha_p}dx^{\alpha_1}\wedge
dx^{\alpha_2}\wedge\dots \wedge dx^{\alpha_p}\quad 0\leq p\leq n\eqno(3.1)
$$
where the local basis obeys the condition of exterior multiplication
$$
\begin{array}{l}
dx^{\alpha}\wedge dx^{\alpha}=0\\
dx^{\alpha}\wedge dx^{\beta}=-dx^{\beta}\wedge dx^{\alpha}\quad
\alpha\ne \beta
\end{array}
$$
(the summation over repeated indices is implied).

But the evolutionary form differential cannot be written similarly to that 
presented for exterior differential forms (see formula (1.3)). In the 
evolutionary form differential there appears an additional term connected with 
the fact that the basis of the form changes. For the differential forms defined 
on the manifold with unclosed metric form one has 
$d(dx^{\alpha_1}dx^{\alpha_2}\dots
dx^{\alpha_p})\neq 0$.  
For this reason the differential of the evolutionary form $\omega^p$ can be 
written as 
$$
d\omega^p{=}\!\sum_{\alpha_1\dots\alpha_p}\!da_{\alpha_1\dots\alpha_p}dx^{\alpha_1}dx^{\alpha_2}\dots
dx^{\alpha_p}{+}\!\sum_{\alpha_1\dots\alpha_p}\!a_{\alpha_1\dots\alpha_p}d(dx^{\alpha_1}dx^{\alpha_2}\dots
dx^{\alpha_p})\eqno(3.2)
$$
where the second term is connected with the differential of the basis. That
is expressed in terms of the metric form commutator. For the manifold with
a closed metric form this term vanishes.

For example, let us consider the first-degree form
$\omega=a_\alpha dx^\alpha$. The differential of this form can
be written as $d\omega=K_{\alpha\beta}dx^\alpha dx^\beta$, where
$K_{\alpha\beta}=a_{\beta;\alpha}-a_{\alpha;\beta}$ are
the components of the commutator of the form $\omega$, and
$a_{\beta;\alpha}$, $a_{\alpha;\beta}$ are the covariant
derivatives. If we express the covariant derivatives in terms of
the connectedness (if it is possible), then they can be written
as $a_{\beta;\alpha}=\partial a_\beta/\partial
x^\alpha+\Gamma^\sigma_{\beta\alpha}a_\sigma$, where the first
term results from differentiating the form coefficients, and the
second term results from differentiating the basis. (In the
Euclidean space the covariant derivatives coincide with ordinary ones
since in this case the derivatives of the basis vanish). If
we substitute the expressions for covariant derivatives into the
formula for the commutator components, then we obtain the following expression
for the commutator components of the form $\omega$:
$$
K_{\alpha\beta}=\left(\frac{\partial a_\beta}{\partial
x^\alpha}-\frac{\partial a_\alpha}{\partial
x^\beta}\right)+(\Gamma^\sigma_{\beta\alpha}-
\Gamma^\sigma_{\alpha\beta})a_\sigma\eqno(3.3)
$$
Here the expressions
$(\Gamma^\sigma_{\beta\alpha}-\Gamma^\sigma_{\alpha\beta})$
entered into the second term are just the components of
commutator of the first-degree metric form.

That is, the corresponding
metric form commutator will enter into the differential form commutator.

Thus, the differentials and, correspondingly, the commutators of exterior and
evolutionary forms are of different types.

\subsection*{Nonclosure of evolutionary differential forms}
The evolutionary differential form commutator, in contrast to that of the 
exterior one, cannot be equal to zero because it involves the metric form 
commutator being nonzero. 
This means that the evolutionary form differential is nonzero. Hence, the 
evolutionary differential form, in contrast to the case of the exterior form,  
cannot be closed.  

Since the evolutionary differential forms are unclosed, the
mathematical apparatus of evolutionary differential forms does not seem to
possess any possibilities connected with the algebraic, group, invariant
and other properties of closed exterior differential forms. However, the
mathematical apparatus of evolutionary forms proves to be significantly
wider due to the fact that evolutionary differential
forms can generate closed exterior differential forms. The mathematical
apparatus of evolutionary differential forms includes, as it will be shown
below, some new unconventional elements.

\subsection*{Nonidentical relations of evolutionary differential forms}
Above it was shown that the identical relations lie at the
basis of the mathematical apparatus of exterior differential forms.

In contrast to this, nonidentical relations lie at the basis of the
mathematical apparatus of evolutionary differential forms.

The nonidentical relation is a relation between a closed exterior differential 
form, which is a differential and is a measurable quantity, and 
an evolutionary form that is an unmeasurable quantity.

Nonidentical relations of such type appear in descriptions of the physical
processes. They may be written as
$$
d\psi \,=\,\omega^p \eqno(3.4)
$$
Here $\omega^p$ is the $p$-degree evolutionary form that is
nonintegrable, $\psi$ is some form of degree $(p-1)$, and
the differential $d\psi$ is a closed form of degree $p$.

This relation is an evolutionary relation as it involves an
evolutionary form.

In the left-hand side of this relation it stands the form differential,
i.e. a closed form that is an invariant object. In the right-hand
side it stands the nonintegrable unclosed form that is not an invariant object.
Such a relation cannot be identical.

One can see a difference of relations for exterior forms and evolutionary ones.
In the right-hand side of identical relation (1.11)
it stands a closed form, whereas the form in the right-hand side of
nonidentical relation (3.4) is an unclosed one. 

{\it How is this relation obtained?}
 
Let us consider this by the example of the first degree differential forms. 
A differential of the function of more than one variables can be an example 
of the first degree form. In this
case the function itself is the exterior form of zero degree.  The state 
function that specifies the state of a material system can serve as
an example of such function. When the physical processes in a material system 
are being described, the state function may
be unknown, but its derivatives may be known. The values of
the function derivatives may be equal to some expressions that
are obtained from the description of a real physical process. And
it is necessary to find the state function. 

Assume that $\psi$ is the desired state function that depends on the 
variables $x^\alpha$ and also assume that its derivatives in
various directions are known and equal to some quantities
$a_\alpha$, namely:
$$
\frac{\partial\psi}{\partial x^\alpha}=a_\alpha\eqno(3.5)
$$
Let us set up the differential expression
$(\partial\psi/\partial x^\alpha)dx^\alpha$ (here the summation
over repeated indices is implied). This differential expression
is equal to
$$
\frac{\partial\psi}{\partial x^\alpha}dx^\alpha=a_\alpha dx^\alpha\eqno(3.6)
$$
Here the left-hand side of the expression is a differential of
the function $d\psi$, and in the right-hand
side it stands the  differential form $\omega=a_\alpha
dx^\alpha$. Relation (3.6) can be written as 
$$
d\psi=\omega\eqno(3.7)
$$
It is evident that relation (3.7) is of the same type as
(3.4) under the condition that the differential form degrees 
are equal to 1. 

This relation is nonidentical because the differential form $\omega $ 
is an unclosed differential form. The commutator of this form is nonzero 
since the expressions $a_\alpha $ for the derivatives 
$(\partial\psi/\partial x^\alpha)$ are nonconjugated quantities. 
They are obtained from the description 
of an actual physical process and are unmeasurable quantities.

One can come to relation (3.7) by means of analyzing the integrability 
of the partial differential equation. An equation is integrable 
if it can be reduced to the form $d\psi=dU$. However it
appears that, if the equation is not subjected to an additional
condition (the integrability condition), it is reduced to the
form (3.7), where $\omega$ is an unclosed form and it cannot be
expressed as a differential. The first principle of thermodynamics is 
an example of nonidentical relation. 

\subsection*{How to work with nonidentical relation?}
While investigating real 
physical processes one often faces the relations that are nonidentical. 
But it is commonly believed that only identical relations can have any 
physical meaning. For this reason one immediately attempts to impose 
a condition onto the nonidentical relation under which this relation
becomes identical, and it is considered only that can satisfy the 
additional conditions. And all remaining is rejected. 
It is not taken into account that a nonidentical relation is often
obtained from a description of some physical process and it has 
physical meaning at every stage of the physical process rather
than at the stage when the additional conditions are
satisfied. In essence the physical process does not considered completely. 

This approach does not solve the evolutionary problem. 

Below we present the evolutionary approach to
the investigation of the nonidentical relation. 

At this point it should be emphasized that the nonidentity of
the evolutionary relation does not mean the imperfect accuracy
of the mathematical description of a physical process. The
nonidentical relations are indicative of specific features of
the physical process development. 

\subsection*{Selfvariation of the evolutionary nonidentical relation}
The evolutionary nonidentical relation is selfvarying,
because, firstly, it is nonidentical, namely, it contains
two objects one of which appears to be unmeasurable, and,
secondly, it is an evolutionary relation, namely, the variation of
any object of the relation in some process leads to variation of
another object and, in turn, the variation of the latter leads to
variation of the former. Since one of the objects is an 
unmeasurable quantity, the other cannot be compared
with the first one, and hence, the process of mutual variation
cannot stop. This process is governed by the evolutionary form commutator. 

Varying the evolutionary form coefficients leads to varying the first 
term of the commutator (see (3.3)). In accordance  with this
variation it varies the second term, that is, the metric form of
the manifold varies. Since the metric form commutators
specifies the manifold differential characteristics that are 
connected with the manifold deformation (for example, the commutator of the 
zero degree metric form specifies the bend, that of
second degree specifies various types of rotation, that of the
third degree specifies the curvature), then it points to the manifold deformation. 
This means that it varies the evolutionary form basis. In turn, it leads to 
variation of the evolutionary form, and the process of intervariation of the 
evolutionary form and the basis is repeated. The processes of variation of 
the evolutionary form and the basis are governed by the evolutionary form 
commutator and it is realized according to the evolutionary relation.

Selfvariation of the evolutionary relation goes on 
by exchange between the evolutionary form coefficients and the manifold 
characteristics. This is an exchange between physical quantities and space-time 
characteristics. (This is an exchange between quantities of different nature).
   
The process of the evolutionary relation selfvariation cannot come to an end. 
This is indicated by the fact that both the evolutionary form commutator 
and the evolutionary relation involve unmeasurable quantities. 

The question arises  whether an identical relation can be obtained from this 
nonidentical selfvarying relation, which would allow determination of the 
desired differential form under the differential sign.

It appears that it is possible under the degenerate transform.

\subsection*{Degenerate transforms.}
To obtain an identical relation from the evolutionary nonidentical relation,  
it is necessary that a closed exterior differential form should be derived 
from the evolutionary differential form that is included into evolutionary  
relation.   

However, as it was shown above, the evolutionary form cannot be a closed form. 
For this reason the the transition from an evolutionary form is possible only 
to an {\it inexact} closed exterior form that is defined on pseudostructure. 

To the pseudostructure there corresponds a closed dual form (whose differential 
vanishes). For this reason a transition from an evolutionary form to a closed 
exterior form proceeds only when the conditions of vanishing the dual form 
differential are realized, in other words, when the the metric form 
differential or commutator becomes equal to zero. 

The conditions of vanishing the dual form differential (additional conditions) 
determine the closed metric form and thereby specify the pseudostructure 
(the dual form).
 
In this case the closed exterior ({\it inexact}) form is formed. 
The evolutionary form commutator and, correspondingly, the evolutionary form 
differential vanish on the pseudostructure. 
 
At this point it should be emphasized that the commutator, that equals zero, 
(and the differential) is an interior one. This commutator is defined only 
on the pseudostructure, the total commutator being nonzero. The evolutionary 
form remains to be unclosed.   

Since the evolutionary form differential is nonzero, whereas the closed exterior 
form differential is zero, then a passage from the evolutionary form to the 
closed exterior form is allowed only under {\it degenerate transform}.

The conditions of vanishing the dual form differential (an additional condition) 
are the conditions of degenerate transform. 

Additional conditions can be realized, for example, if it will appear any 
symmetries of the evolutionary form coefficients  or its commutator. This can 
happen under selfvariation of the nonidentical relation. (Whyle describing  
material system such additional conditions are related, for example, to 
additional degrees of freedom). 

Mathematically to the conditions of degenerate transform there corresponds 
a requirement that some functional expressions become equal 
to zero. Such functional expressions are Jacobians, determinants, the Poisson brackets, 
residues, and others. 

\subsection*{Obtaining an identical relation from a nonidentical one}
Assume that the nonidentical relation has the form of (3.4) and from the 
evolutionary form $\omega^p$ the differential form closed 
on pseudostructure was obtained. Since the closed exterior form is a 
differential, in this case 
the evolutionary relation (3.4) becomes identical on the pseudostructure. From 
this relation one can find the desired differential $d_\pi\psi$ that is 
equal to the closed exterior differential form derived. One can obtain the 
desired form $\psi$ from this differential. 

It can be shown that all identical relations of the exterior differential 
form theory are obtained from nonidentical relations (that contain the 
evolutionary forms) by applying the degenerate transform.

Under degenerate transform the evolutionary form differential vanishes only
on pseudostructure. The total differential 
of the evolutionary form is nonzero. The evolutionary form tremains unclosed.

\subsection*{Integration of the nonidentical evolutionary relation}
Under degenerate transform from the nonidentical evolutionary relation one 
obtains a relation being identical on pseudostructure. It is just a relation 
that one can integrate
and obtain the relation with the differential forms of less by one degree.

The relation obtained after integration proves to be nonidentical as well. 
 
The obtained nonidentical relation of degree $(p-1)$ can be integrated once
again if the corresponding degenerate transform is realized and the identical
relation is formed.

By sequential integrating the evolutionary relation of degree $p$ (in the case 
of realization of the corresponding degenerate transforms and forming
the identical relation), one can get closed (on the pseudostructure) exterior 
forms of degree $k$, where $k$ ranges from $p$ to $0$. 

In this case one can see that under such integration closed (on the 
pseudostructure) exterior forms, which depend on two parameters, are obtained.
These parameters are the degree of evolutionary form $p$ 
(in the evolutionary relation) and the degree of created closed forms $k$.

In addition to these parameters, another parameter appears, namely, the 
dimension of space.
If the evolutionary relation generates the closed forms
of degrees
$k=p$, $k=p-1$, \dots, $k=0$, to them there correspond the pseudostructures
of dimensions $(N-k)$, where $N$ is the space dimension. 
It is known that to the closed exterior differential forms of degree $k$
there correspond skew-symmetric tensors of rank $k$ and to corresponding
dual forms there do the pseudotensors of rank $(N-k)$, where $N$ is
the space dimensionality. The pseudostructures correspond to such tensors,
but only on the space formed.

\subsection*{The properties of pseudostructures and closed exterior forms.
Forming fields and manifolds}
As mentioned before, the additional conditions, namely, the conditions
of degenerate transform, specify the pseudostructure. But at every stage
of the evolutionary process it is realized only one element of pseudostructure,
namely, a certain minipseudostructure. The additional conditions determine the
direction (a derivative that specifies the minipseudostructure) on which the 
evolutionary form differential vanishes. (However, in this case the total 
differential of the evolutionary form is nonzero). The closed exterior form 
is formed along this direction. 

While varying the evolutionary variable the minipseudostructures form
the pseudostructure.

The example of minipseudoctructure is the wave front. The wave front
is the eikonal surface (the level surface), i.e. the surface with
a conservative quantity.
The direction that specifies the pseudostructure is a connection between
the evolutionary and spatial variables. It gives the rate of changing the
spatial variables. Such a rate is a velocity of the wave front translation.
While its translation the wave front forms the pseudostructure.

The manifolds with closed metric forms are formed by pseudostructures.
They are obtained from manifolds with unclosed metric forms. In this case
the initial manifold (on which the evolutionary form is defined) and
the formed manifold with closed metric forms (on which the closed exterior form
is defined) are different spatial objects.

It takes place a transition from the initial manifold with unclosed metric form to
the pseudostructure, namely, to the created manifold with closed metric forms.
Mathematically this transition (degenerate transform) proceeds as a transition
from one frame of reference to another, nonequivalent, frame of reference.

The pseudostructures, on which the closed {\it inexact} forms are defined,
form the pseudomanifolds (integral surfaces, pseudo-Riemann and
pseudo-Euclidean spaces are the examples of such manifolds). In this process
the dimensions of formed manifolds are connected with the evolutionary
form degree (This will be shown in sections 5 and 6).

To the transition from pseudomanifolds to metric manifolds there corresponds
the transition from closed {\it inexact} differential forms to {\it exact} 
exterior differential forms. (Euclidean and Riemann spaces are the examples 
of metric manifolds).

Here it is to be noted that the examples of pseudometric spaces are the potential
surfaces (the surfaces of a simple layer, a double layer and so on). In these
cases a type of the potential surfaces is connected with the above listed
parameters.

The fields of conservative quantities (the physical fields) 
are formed from closed exterior forms
at the same time when the manifolds are created from the pseudoctructures.
The specific feature of the manifolds with closed metric
forms that have been formed is that they can carry some information. 

Since the closed metric form is dual with respect to some closed exterior
differential form, the metric forms cannot become closed by themselves,
independently of the exterior differential form. This proves that the manifolds with
closed metric forms are connected with the closed exterior differential forms.
This indicates that the fields of conservative quantities are formed from 
closed exterior forms at the same time when the manifolds are created from the 
pseudoctructures. The specific feature of the manifolds with closed metric
forms that have been formed is that they can carry some information.

That is, the closed exterior differential forms and manifolds, on which they
are defined, are mutually connected objects. On the one hand, this shows 
duality of these two objects (the pseudostructure and the closed inexact 
exterior form), and, on the  other hand, this means that these objects 
constitute a unified whole. This whole is a new conjugated object. 
This is an example of the differential and geometrical structure (G-Structure). 
Such binary object can be named as Bi-Structure. 

\subsection*{Characteristics of the Bi-Structure}
Since the closed exterior differential form that corresponds to the arisen
Bi-Structure was obtained from the nonidentical relation that involves the
evolutionary form, it is evident that the Bi-Structure characteristics
must be connected with those of the evolutionary form and of the manifold
on which this form is defined, with the conditions of degenerate transform
and with the values of commutators of the evolutionary form and the
manifold metric form.

While describing the mechanism of Bi-Structure origination one can see that
at the instant the Bi-Structure originates there appear the following typical
functional expressions and quantities: (1) the condition of degenerate
transform, i.e. vanishing of the interior commutator of the metric form;
(2) vanishing of the interior commutator
of the evolutionary form; (3) the value of the nonzero total commutator of
the evolutionary form that involves two terms, namely,
the first term is composed of the derivatives of the evolutionary form
coefficients, and the second term is composed of the derivatives of the
coefficients of the dual form that is connected with the manifold (here
we deal with a value that the evolutionary form commutator assumes at the
instant of Bi-Structure origination).
They determine the following characteristics of the Bi-Structure. The conditions
of degenerate transform, as it was said before, determine the pseudostructures.
The first term of the evolutionary form commutator determines the value of
the discrete change (the quantum), which the quantity conserved on the
pseudostructure undergoes at the transition from one pseudostructure to another.
The second term of the evolutionary form commutator specifies a characteristic
that fixes the character of the initial manifold deformation that took place
before the Bi-Structure arose. (Spin is such an example).

A discrete (quantum) change of a quantity proceeds in the direction
that is normal (more exactly, transverse) to the pseudostructure. Jumps of the
derivatives normal to the potential surfaces are examples of such changes.

Bi-Structure may carry a physical meaning. Such bynary objects are the physical structures 
from which the physical fields are formed. (About this it will be written below).
\subsection*{Transition from nonconjugated operators to
conjugated operators}
The mathematical apparatus of the evolutionary differential forms reveals
the process of
conjugating the operators. A mutual variations of the evolutionary form
coefficients (which have the algebraic nature) and the manifold characteristics
(which have the geometric nature) and the
realization of the degenerate transform as well is the mechanism
of this process. Hence one can see that the process of
conjugating the operators is a mutual exchange between the physical
and spatial quantities (between the quantities of different nature)
and the degenerate transform under additional conditions. 

The evolutionary differential form is an unclosed form, that is, it is the
form whose differential is not equal to zero. The differential of the
exterior differential form equals zero. To the closed exterior
form there correspond conjugated operators, whereas to the evolutionary form
there correspond nonconjugated operators. The transition from
the evolutionary form to the closed exterior form is that from
nonconjugated operators to conjugated ones. This is expressed
mathematically as the  transition from a nonzero differential (the evolutionary 
form differential is nonzero) to a differential that equals zero (the closed 
exterior form differential equals zero), and this is effected as a transition 
from one coordinate system to another (nonequivalent) coordinate system.  

%\bigskip
In conclusion of this section it is essential to emphasize the following.
By comparing the exterior and evolutionary forms one can see that they possess
the opposite properties. Yet it was shown that the
evolutionary differential forms generate the closed exterior
differential forms. This elucidates that the exterior and
evolutionary differential forms are the unified whole. 

\section{Role of exterior and evolutionary forms in mathematical physics}
The role of exterior and evolutionary  differential forms in mathematical 
physics is due to the fact that the mathematical apparatus of these forms make 
it possible to keep track of the conjugacy of the equations that describe 
physical processes. The functional properties of the solutions of differential
equations are just depend on whether or not the conjugacy conditions are
satisfied. If these equations 
(or derivatives with respect to different variables) be not conjugated, 
then the solutions to corresponding equations prove to be noninvariant: 
they are functionals rather then functions. The realization of the conditions
(while varying variables), under which the equations become conjugated ones, 
leads to that the relevant solution becomes invariant.  
The transition to the invariant solution, which can be obtained only with the 
help of evolutionary forms, describes the various evolutionary transition. 

In field theory only invariant solutions (to which the closed exterior forms 
correspond), and the evolutionary process is thereby not considered.

An investigation of differential equations, which decribe physical processes, 
by means of the exterior and evolutionary differential forms allows to 
understand a mechanism of the evolutionary processes and origination of 
physical structures (of which the physical fields  are formed) and explains 
such processes as fluctuations, pulsations, creation of massless particles, 
emergence of waves, vortices and so on. 

\{It can be shown that the exterior and evolutionary differential forms allow 
to investigate a conjugacy of any differential equations. Ant this lies at 
the basis of the qualitive theory of differential equations. The conjugacy of 
equations points to integrability of equations.\}

The basic idea of the investigation of the conjugacy of differential 
equations can be clarified by the example of the first-order
partial differential equation.

Let
$$ 
F(x^i,\,u,\,p_i)=0,\quad p_i\,=\,\partial u/\partial x^i \eqno(4.1)
$$
be the partial differential equation of the first order. Let us consider the
functional relation
$$ 
du\,=\,\theta\eqno(4.2)
$$
where $\theta\,=\,p_i\,dx^i$ (the summation over repeated indices is implied).
Here $\theta\,=\,p_i\,dx^i$ is
the differential form of the first degree.

The left-hand side of this relation involves the differential, and
the right-hand side includes the differential form  $\theta\,=\,p_i\,dx^i$. 

In the general case, from equation (4.1) it does not follow (explicitly)
that the derivatives $p_i\,=\,\partial u/\partial x^i$ that obey
to the equation (and given boundary or initial conditions of the problem)
make up a differential. 
The commutator of the differential form $\theta$ defined as
$K_{ij}=\partial p_j/\partial x^i-\partial p_i/\partial x^j$ is
not equal to zero. The form $\theta\,=\,p_i\,dx^i$ proves to be unclosed and is not a
differential like the left-hand side of relation (4.2). The functional
relation (4.2) appears to be nonidentical: the left-hand side of this relation
is a differential, but the right-hand side is not a differential. The form 
$\theta\,=\,p_i\,dx^i$ is an example of the evolutionary form, whereas 
functional relation (4.2) is an example of the evolutionary relation. 

The nonidentity of  functional relation (4.2) points to a fact that without 
additional conditions the derivatives of the initial equation do not make 
up a differential (they are nonconjugated). This means that the corresponding
solution to the differential equation $u$ will not be a function of $x^i$.
It will depend on the commutator of the form $\theta $, that is, it will be
a functional.

To obtain the solution that is the function (i.e., the derivatives of this
solution format a differential), it is necessary to add a closure condition
for the form $\theta\,=\,p_idx^i$ and for the dual form (in the present
case the functional $F$ plays a role of the form dual to $\theta $) [2]:
$$\cases {dF(x^i,\,u,\,p_i)\,=\,0\cr
d(p_i\,dx^i)\,=\,0\cr}\eqno(4.3)$$
If we expand the differentials, we get a set of homogeneous equations
with respect to $dx^i$ and $dp_i$ (in the $2n$-dimensional space -- initial
and tangential):
$$\cases {\displaystyle \left ({{\partial F}\over {\partial x^i}}\,+\,
{{\partial F}\over {\partial u}}\,p_i\right )\,dx^i\,+\,
{{\partial F}\over {\partial p_i}}\,dp_i \,=\,0\cr
dp_i\,dx^i\,-\,dx^i\,dp_i\,=\,0\cr} \eqno(4.4)$$
The solvability conditions for this set (vanishing of the determinant
composed of the coefficients at $dx^i$, $dp_i$) have the form:
$$
{{dx^i}\over {\partial F/\partial p_i}}\,=\,{{-dp_i}\over {\partial F/\partial x^i+p_i\partial F/\partial u}} \eqno (4.5)
$$
These conditions determine an integrating direction, namely, a pseudostructure,
on which the form $\theta \,=\,p_i\,dx^i$ turns out to be closed one,
i.e. it becomes a differential, and from relation (4.2) the identical relation
is produced. If conditions (4.5), that may be called conjugacy conditions 
(integrability conditions), are satisfied, the derivatives constitute
a differential $\delta u\,=\,p_idx^i\,=\,du$ (on the pseudostructure), and the
solution becomes a function. Just such solutions, namely, functions on the 
pseudostructures formed by the integrating directions, are the so-called 
generalized solutions [19]. The derivatives of the generalized solution form 
the exterior form that is closed on the pseudostructure.

If conditions (4.5) are not satisfied, that is, the derivatives do not
form a differential, the solution that corresponds to such derivatives
will depend on the differential form commutator formatted by derivatives.
That means that the solution is a functional rather then a function. 

Since the functions that are the generalized solutions
are defined only on the pseudostructures, they have discontinuities in
derivatives in the directions that are transverse to the pseudostructures.
The order of derivatives with discontinuities
is equal to  the exterior form degree. If the form of zero
degree is involved in the functional relation, then the function itself,
being a generalized solution, will have discontinuities.

If we find the characteristics of equation (4.1), it appears that
conditions (4.5) are the equations for the characteristics [14].
That is, the characteristics are the examples of
the pseudostructures on which the derivatives of the differential equation
form the closed forms and the solutions prove to be the functions (generalized
solutions). (The characteristic manifolds of equation (4.1) are the
pseudostructures $\pi$ on which the form $\theta =p_idx^i$ becomes a closed
form: $\theta _{\pi}=d u_{\pi}$).

Here it is worth noting that the coordinates of the equations for
characteristics are not identical to the independent coordinates of the
initial space on which equation (4.1) is defined. The transition from
the initial space to the characteristic manifold appears to be a
{\it degenerate} transform, namely,
the determinant of the set of equations (4.4) becomes zero. The derivatives
of  equation (4.1) are transformed from the tangent space to the cotangent one.

Similar functional properties have the solutions to all differential equations 
and a system of partial differential equations. 

Differential equations of mathematical physics, which describe physical 
processes, are those on which the invariance requirements (conjugacy 
conditions) are not inposed. Therefore they can have solutions that are 
functionals (their derivatives do not make up a differential). The solutions 
prove to be exact (generalized, not functionals) only under realization of
the additional requirements, namely, the conditions of degenerate transforms:
vanishing the determinants, Jacobians and so on, that define the integral
surfaces. The characteristic manifolds, the envelopes of 
characteristics, singular points, potentials of simple and double layers, 
residues and others are the examples of such surfaces.

Since the closed forms correspond to the physical structures that
form the physical fields, it is evident that only the equations with
additional conditions (the integrability conditions) can be the equations
of the field theory.

Let us consider how such an equation can be obtained and to do so we
return to equation (4.1).

Assume that it does not explicitly
depend on $u$ and it is solved with respect to some variable, for example
$t$, that is, it has the form of
$${{\partial u}\over {\partial t}}\,+\,E(t,\,x^j,\,p_j)\,=\,0, \quad p_j\,=\,{{\partial u}\over {\partial x^j}}\eqno(4.6)
$$
Then integrability conditions (4.5)
(the closure conditions of the differential form
$\theta =p_idx^i$  and the corresponding dual form) can be written as
(in this case $\partial F/\partial p_1=1$)
$${{dx^j}\over {dt}}\,=\,{{\partial E}\over {\partial p_j}}, \quad
{{dp_j}\over {dt}}\,=\,-{{\partial E}\over {\partial x^j}}\eqno(4.7)
$$

These are the characteristic relations (equations of characteristics) for 
equation (4.6). As it is well known, the canonical relations have just 
such a form.

Equation (4.6) provided with the supplementary conditions, namely, 
the canonical relations (4.7), is called the Hamilton-Jacobi equation [14]. 
The equations of field theory 
$${{\partial s}\over {\partial t}}+H \left(t,\,q_j,\,{{\partial s}\over {\partial q_j}}
\right )\,=\,0,\quad
{{\partial s}\over {\partial q_j}}\,=\,p_j \eqno(4.8)$$ 
belong to this type.
Here $s$ is the field function for the action functional $S\,=\,\int\,L\,dt$.
$L$ is the Lagrange function, $H$ is the Hamilton function:
$H(t,\,q_j,\,p_j)\,=\,p_j\,\dot q_j-L$, $p_j\,=\,\partial L/\partial \dot q_j$.
The closed form $ds\,=\,-H\,dt\,+\,p_j\,dq_j$ (the
Poincare invariant) corresponds to equation (4.8).

Here the degenerate transform is a transition from the Lagrange
function to the Hamilton function. The equation for the Lagrange function, that
is the Euler variational equation, was obtained from the condition $\delta S\,=\,0$,
where $S$ is the action
functional. In the real case, when forces are nonpotential or couplings are
nonholonomic, the quantity $\delta S$ is not a closed form, that is, $d\,\delta S\,\neq \,0$.
But the Hamilton function is obtained from the condition $d\,\delta S\,=\,0$
which is the closure condition for the form $\delta S$. The transition from the
Lagrange function $L$ to the Hamilton function $H$ (the transition from
variables $q_j,\,\dot q_j$ to variables $q_j,\,p_j=\partial L/\partial \dot q_j$)
is the transition from the tangent space, where the form is unclosed, to
the cotangent space with a closed form. One can see that this transition is
a degenerate one.

In quantum mechanics (where to the coordinates $q_j$, $p_j$ the operators are
assigned) the Schr\H{o}dinger equation [20] serves as an analog to equation 
(4.8), and the Heisenberg equation serves as an analog to the relevant equation
for the canonical relation integral. Whereas the closed
exterior differential form of zero degree (the analog to the Poincare
invariant) corresponds to the Schr\H{o}dinger equation, the closed dual form
corresponds to the Heisenberg equation.

The invariant field theories used only nondegenerate transformations that
conserve the differential.
By the example of the canonical relations it is possible to show that
nondegenerate and degenerate transformations are connected. The canonical
relations in the invariant field
theory correspond to nondegenerate tangent transformations.
At the same time, the canonical relations coincide  with the characteristic
relation for  equation (4.8), which
the degenerate transformations correspond to. The degenerate transformation is
a transition from the tangent space ($q_j,\,\dot q_j)$) to the
cotangent (characteristic)
manifold ($q_j,\,p_j$). On the other hand, the nondegenerate
transformation is a transition
from one characteristic manifold ($q_j,\,p_j$) to the other characteristic
manifold ($Q_j,\,P_j$). 

The investigations conducted show that one must separate out two types 
of differential equations, namely, differential equations, that describe 
physical processes, and the invariant equations of the field theory, that describe 
physical structures forming physical fields. As the investigations conducted 
show these equations are connected between themselves. Invariant equations are 
obtained from equations that describe physical processes on which there are 
imposed additional conditions, namely, the 
conditions of existence of closed forms (invariants). 

A connection of two types of equations of mathematical physics elucidates a 
relation between physical fields and material media. 
Differential equations that describe 
physical processes are equations of the conservation laws for material media 
whereas invariant equations of the field theory are equations 
(more presicely, relations) of the conservation laws for physical fields. 
Below with it will be 
shown that material media generate physical fields. 
And this process is controlled by the conservation laws. 

\section{Conservation laws}
One has to distinguish two types of the conservation laws that for
convenience can be called the exact conservation laws and the balance ones.

\subsection*{Exact conservation laws}
{\it The exact conservation laws are those that state the existence of
conservative physical quantities or objects. The exact
conservation laws are related to physical fields}. \{The physical fields [21] are
a special form of the substance, they are carriers of various interactions
such as electromagnetic, gravitational, wave, nuclear and other kinds of
interactions.\}

The closed exterior differential forms correspond to the exact conservation laws.
Indeed, from the closure conditions of the exterior differential form
(see Section 1, formulas (1.4), (1.8), (1.9)) it is evident that a closed
differential form is a conservative quantity. In this case the closed inexact
differential form and the corresponding dual form describe a conservative
object, namely, there is a conservative quantity only on some pseudostructure
$\pi $. From this one can see that the closed exterior differential form can
correspond to the exact conservation law.

The closure conditions for the exterior differential
form ($d_{\pi }\,\theta ^p\,=\,0$)
and the dual form ($d_{\pi }\,^*\theta ^p\,=\,0$) are
mathematical expressions of the exact conservation law.

In section 1 and 3 it was mentioned that the pseudostructure
and the closed exterior form defined on the pseudostructure make up a binary
structure (Bi-Structure). It is evident that such a structure does correspond
to the exact conservation law.

It is such structures (pseudostructures with a conservative physical quantity)
that corresponding to exact conservation law that are the physical structures
from which physical
fields are formed. The problem of how these structures arise and how
physical fields are formed will be discussed below.

Equations for the physical structures ($d_{\pi }\,\theta ^p\,=\,0$,
$d_{\pi }\,^*\theta ^p\,=\,0$) turn out to coincide with the mathematical
expression for the exact conservation law.

The mathematical expression for the exact conservation law and its connection
with physical fields can be schematically written in the following manner
$$
\def\\{\vphantom{d_\pi}}
\cases{d_\pi \theta^p=0\cr d_\pi {}^{*\mskip-2mu}\theta^p=0\cr}\quad
\mapsto\quad
\cases{\\\theta^p\cr \\{}^{*\mskip-2mu}\theta^p\cr}\quad\hbox{---}\quad
\hbox{physical structures}\quad\mapsto\quad\hbox{physical fields}
$$

It is seen that the exact conservation law is that for physical fields.

It should be emphasized that the closed {\it inexact} forms correspond to
the physical structures that form physical fields. The {\it exact} forms
correspond to the material system elements. (About this it will be said below).

It can be shown that the field theories, i.e. the theories
that describe physical fields, are based on the invariant and metric
properties of the closed exterior differential and dual forms
that correspond to exact conservation laws.

\subsection*{Balance conservation laws}
{\it The balance conservation laws are those that establish the balance between
the variation of a physical quantity and the corresponding external action.
These are the conservation laws for the material systems (material media)}.

\{A material 
system is a variety of elements which have internal structure and interact 
to one another. As examples of material systems it may be thermodynamic, 
gas dynamical, cosmic systems, systems of elementary particles  
and others. The physical vacuum in its properties may be regarded as an analogue 
of the material system that generates some physical fields. Any material media 
are such material systems. Examples of elements that constitute the material system 
are electrons, protons, neutrons, atoms, fluid particles, cosmic objects and 
others\}.

The balance conservation laws are the conservation laws for energy, linear
momentum, angular momentum, and mass.

In the integral form the balance conservation laws express the following [22]:
a change of a physical quantity in an elementary volume over a time interval
is counterbalanced by the flux of a certain quantity through the boundary surface
and by the action of sources. Under transition to the differential expression
the fluxes are changed by divergences.

The equations of the balance conservation laws are differential (or integral)
equations that describe a variation
of functions corresponding to physical quantities [16,22-24].

It appears that, even without knowledge of the concrete form
of these equations, with the help of the differential forms one can see
specific features of these equations that elucidate the properties of
the balance conservation laws. To do so it is necessary to study the conjugacy
(consistency) of these equations.

Equations are conjugate if they can be contracted into identical
relations for the differential, i.e. for a closed form.

Let us analyze the equations
that describe the balance conservation laws for energy and linear momentum.

We introduce two frames of reference: the first is an inertial one
(this frame of reference is not connected with the material system), and
the second is an accompanying
one (this system is connected with the manifold built by
the trajectories of the material system elements). The energy equation
in the inertial frame of reference can be reduced to the form:
$$
\frac{D\psi}{Dt}=A_1 \eqno(5.1)
$$
where $D/Dt$ is the total derivative with respect to time, $\psi$ is the
functional
of the state that specifies the material system, $A_1$ is the quantity that
depends on specific features of the system and on external energy actions onto
the system. \{The action functional, entropy,wave function
can be regarded as examples of the functional $\psi $. Thus, the equation
for energy presented in terms of the action functional $S$ has a similar form:
$DS/Dt\,=\,L$, where $\psi \,=\,S$, $A_1\,=\,L$ is the Lagrange function.
In mechanics of continuous media the equation for
energy of an ideal gas can be presented in the form [23]: $Ds/Dt\,=\,0$, where
$s$ is entropy. In this case $\psi \,=\,s$, $A_1\,=\,0$. It is worth noting that
the examples presented show that the action functional and entropy play the
same role.\}

In the accompanying frame of reference the total derivative with respect to
time is transformed into the derivative along the trajectory. Equation (5.1)
is now written in the form
$$
{{\partial \psi }\over {\partial \xi ^1}}\,=\,A_1 \eqno(5.2)
$$
here $\xi^1$ is the coordinate along the trajectory.

In a similar manner, in the
accompanying frame of reference the equation for linear momentum appears
to be reduced to the equation of the form 
$$
{{\partial \psi}\over {\partial \xi^{\nu }}}\,=\,A_{\nu },\quad \nu \,=\,2,\,...\eqno(5.3)
$$
where $\xi ^{\nu }$ are the coordinates in the direction normal to the
trajectory, $A_{\nu }$ are the quantities that depend on the specific
features of the system and external force actions.

Eqs. (5.2), (5.3) can be convoluted into the relation
$$
d\psi\,=\,A_{\mu }\,d\xi ^{\mu },\quad (\mu\,=\,1,\,\nu )\eqno(5.4)
$$
where $d\psi $ is the differential
expression $d\psi\,=\,(\partial \psi /\partial \xi ^{\mu })d\xi ^{\mu }$.

Relation (5.4) can be written as
$$
d\psi \,=\,\omega \eqno(5.5)
$$
here $\omega \,=\,A_{\mu }\,d\xi ^{\mu }$ is the differential form of the
first degree.

Since the balance conservation laws are evolutionary ones, the relation
obtained is also an evolutionary relation.

Relation (5.5) was obtained from the equation of the balance
conservation laws for
energy and linear momentum. In this relation the form $\omega $ is that of the
first degree. If the equations of the balance conservation laws for
angular momentum be added to the equations for energy and linear momentum,
this form in the evolutionary relation will be the form of the second degree.
And in  combination with the equation of the balance conservation law
of mass this form will be the form of degree 3.

Thus, in the general case the evolutionary relation can be written as
$$
d\psi \,=\,\omega^p \eqno(5.6)
$$
where the form degree  $p$ takes the values $p\,=\,0,1,2,3$..
(The evolutionary
relation for $p\,=\,0$ is similar to that in the differential forms, and it was
obtained from the interaction of energy and time.)

In relation (5.5) the form $\psi$ is the form of zero degree. And in relation
(5.6) the form $\psi$ is the form of $(p-1)$ degree.

Let us show that {\it the evolutionary relation  obtained from the equation
of the balance conservation laws proves to be nonidentical}.

To do so we shall analyze relation (5.5).

A relation may be identical one if this is a relation between measurable
(invariant) quantities or between observable (metric) objects, in other
words, between quantities or objects that are comparable.

In the left-hand side of evolutionary relation (5.5) there is a
differential that is a closed form. This form is an invariant
object. Let us show that the differential form
$\omega$ in the right-hand side of relation (5.5) is not an 
invariant object because in real processes this form proves to be unclosed.

For the form to be closed the differential of the form or its commutator
must be equal to zero (the elements of the form differential are equal to
the components of its commutator).

Let us consider the commutator of the
form $\omega \,=\,A_{\mu }d\xi ^{\mu }$.
The components of the commutator of such a form (as it was pointed above) can
be written as follows:
$$
K_{\alpha \beta }\,=\,\left ({{\partial A_{\beta }}\over {\partial \xi ^{\alpha }}}\,-\,
{{\partial A_{\alpha }}\over {\partial \xi ^{\beta }}}\right )\eqno(5.7)
$$
(here the term  connected with the nondifferentiability of the manifold
has not yet been taken into account).

The coefficients $A_{\mu }$ of the form $\omega $ have been obtained either
from the equation of the balance conservation law for energy or from that for
linear momentum. This means that in the first case the coefficients depend
on the energetic action and in the second case they depend on the force action.
In actual processes energetic and force actions have different nature and appear
to be inconsistent. The commutator of the form $\omega $ constructed from
the derivatives of such coefficients is nonzero.
This means that the differential of the form $\omega $
is nonzero as well. Thus, the form $\omega$ proves to be unclosed and is not
a measurable quantity.

This means that the evolutionary relation
involves an unmeasurable term. Such a relation cannot be an identical one.
(In the left-hand side of this relation it stands a differential, whereas
in the right-hand side it stands an unclosed form that is not a differential.)

Hence,  without  knowledge of a particular expression for the form $\omega$,
one can argue that for actual processes the evolutionary relation proves to be
nonidentical because of inconsistency of the external action.

The nonidentity of the evolutionary relation means that equations of the
balance conservation laws turn out to be nonconjugated (thus, if from
the energy equation we obtain the  derivative of $\psi $ in the direction
along the trajectory and from the momentum equation we find the derivative
of $\psi $ in the direction normal to the trajectory and then we to calculate
their mixed derivatives, then
from the condition that the commutator of the form $\omega $
is nonzero it follows that the mixed derivatives prove to be noncommutative).
One  cannot  convolute them into an identical relation and obtain a
differential.

The nonconjugacy of the balance conservation
law equations reflects the properties of the balance conservation laws that
have a governing importance for the evolutionary processes, namely, their
noncommutativity.

\section{Evolutionary processes in material media 
and origination of physical structures}
It was shown above that the evolutionary relation that was obtained from
the balance conservation law equations proves to be nonidentical.
This points to the noncommutativity of the balance conservation laws. By
analyzing the behavior of the nonidentical evolutionary relation one can
understand to what result the noncommutativity of the balance conservation laws 
leads.

Let us consider  evolutionary relation (5.6).

In the left-hand side of the evolutionary relation there is the functional
expression $d\psi$ that determines the state of the material system, and in 
the right-hand side there is the form $\omega^p$ whose coefficients depend on
external actions and the material system characteristics.

The evolutionary relation gives a possibility to
determine either presence or absence of the differential (the closed form).
And this allows us to recognize whether the material system state is 
in equilibrium, in local equilibrium or not in equilibrium. If it is 
possible to determine the differential $d\psi$ (the state differential) from
the evolutionary relation, this indicates that the system is in equilibrium
or locally equilibrium state. And if the differential cannot be determined, 
then this means that the system is in a nonequilibrium state.

It is evident that if the balance conservation laws be commutative,
the evolutionary relation would be identical and from that it would be possible 
to get the differential $d\psi $, this would indicate  that the material system 
is in the equilibrium state.

However, as it has been shown, in real processes the balance conservation laws
are noncommutative. The evolutionary relation is not identical and from this 
relation one cannot get the differential $d\psi $. This means that the system 
state is nonequilibrium. It is evident that a magnitude of internal force, which 
gives rise to the nonequilibrium, is described by the evolutionary form 
commutator. 

\subsection*{Selfvariation of nonequilibrium state of material system. 
(Selfvariation of the evolutionary relation)}
What does the material system nonequilibrium indicated
by the nonidentity of the evolutionary relation results in?

As it was shown in Section 3, if the relation is
evolutionary and nonidentical, it is a selfvarying relation, that is,
a change of one object in the relation leads to a change of the other object,
and in turn a change of the latter leads to a change of the former and so on.
Such a specific feature of
the evolutionary relation explains the particulars of the material
system, namely, {\it the selfvariation of its nonequilibrium
state}.

The selfvariation mechanism of the {\it nonequilibrium}
state of the material system can be understood if analyze 
the properties of the evolutionary form commutator.

The evolutionary form in the evolutionary relation is defined on the
accompanying manifold that for real processes appears to be the deformable
manifold because it is formed simultaneously with a change of the
material system state and depends on the physical processes.
Such a manifold cannot be a manifold with closed metric forms. Hence, as it was 
already noted in Section 3, the term containing the characteristics of the
manifold will be included into the evolutionary form commutator in
addition to the term connected with derivatives of the form
coefficients. The interaction between these terms of
different nature describes a mutual change of
the state of the material system.

The emergence of the second term under deformation of the accompanying 
manifold can only change the commutator and cannot make it zero (because 
the terms of the commutator have different nature). In the
material system the internal force will continue to act even in the absence
of external actions. The further deformation (torsion) of the manifold
will go on. This leads to a change of the metric form commutator, 
produces a change of the evolutionary form and its commutator and so on. 
Such a process is governed by the nonidentical evolutionary relation and, 
in turn, produces a change of the evolutionary relation.

The process of selfvariation of the commutator controlled by the
nonidentical evolutionary relation specifies a
change of the internal force that acts in the material system. This points
to a change of the material system state. But the material system state
remains nonequilibrium in this process because the internal forces
do not vanish due to the evolutionary form commutator remaining nonzero.

At this point it should be emphasized that such selfvariation of the material
system state proceeds under the action of internal (rather than external)
forces. That will go on even in the absence of external forces. That is,
the selfvariation of the
nonequilibrium state of the material system takes place.

It is the selfvariation of the nonidentical evolutionary relation obtained from 
the balance conservation laws for material systems that describes the
selfvaration of the nonequilibrium state of material system.

Here it should be noted that in a real physical process the internal forces
can be increased (due to the selfvariation of the nonequilibrium state of the
material system).  This can lead to the development of instability in the 
material system [25]. 
\{For example, this was pointed out in the works by Prigogine
[26]. ``The excess entropy" in his
works is analogous to the commutator of a nonintegrable form for the
thermodynamic system.  ``Production of excess entropy" leads to the
development of instability\}.

\subsection*{Transition of the material system into a locally equilibrium
state. (Degenerate transform)}
Now the question arises whether the material system can got rid of
the internal force and transfer into the equilibrium state?

The state of material system is equilibrium one if the 
state differential exists.

The state differential is included into the nonidentical evolutionary relation.
But one cannot get the state differential from {\it nonidentical} relation. 
This points to that the material system cannot come to the equilibrium state.  
However, as it was shown in Section 3, under degenerate transform the identical 
relation can be obtained from the nonidentical evolutionary relation. From such 
a relation one can obtain the state differential. But to this differential there 
corresponds the {\it inexact } closed exterior form. this means that the state 
differential acts only locally. The availability of such state diferential 
points not to the equilibriun state but to the locally equilibrium state of 
material system. Under this condition the total state of material system 
remains to be nonequilibrium.

How does the transition of the material system from the nonequilibrium state to a
locally equilibrium one go on?

Let us consider nonidentical evolutionary relation (5.6).

As it was shown in Section 3, from the nonidentical evolutionary relation 
it can be obtained the identical relation if from the evolutionary unclosed 
relation the inexact closed exterior form can be obtained.

Since the evolutionary differential form $\omega^p$ is unclosed, the 
commutator, and hence the differential, of this form is nonzero. That is,
$$
d\omega^p\ne 0\eqno(6.1)
$$
The unclosed inexact exterior form, which can be obtained from the 
evolutionary form, must satisfy the closure conditions:
$$
\left\{
\begin{array}{l}
d_\pi \omega^p=0\\
d_\pi{}^*\omega^p=0
\end{array}
\right.\eqno(6.2)
$$
That is, the differential of this form must be equal to zero. It is evident 
that a passage from condition (6.1) to condition (6.2) is allowed  only as 
the degenerate transform. If such a transform takes place, on the 
pseudostructure $\pi$ evolutionary relation (5.6) transforms into the relation
$$
d_\pi\psi=\omega_\pi^p\eqno(6.3)
$$
which proves to be the identical relation. Indeed, since the form
$\omega_\pi^p$ is a closed one, on the pseudostructure it turns
out to be a differential of some differential form. In other words,
this form can be written as $\omega_\pi^p=d_\pi\theta$. Relation (6.3)
is now written as
$$
d_\pi\psi=d_\pi\theta
$$
There are differentials in the left-hand and right-hand sides of
this relation. This means that the relation is an identical one.

It is from such a relation that one can find the state differential 
$d_\pi\psi$. 

The emergence of the differential $d_\pi\psi$ indicates that the material
system changes into the locally equilibrium state.
{\it But in this case the total state of the material system turns out to
be nonequilibrium.}

Here the following detail should be taken into account. The evolutionary 
differential form is defined on the accompanying manifold. The evolutionary 
relation has been obtained under the condition that the coordinate frame 
is tied to the accompanying manifold. This coordinate system is not an 
inertial or locally inertial system because the metric forms of the 
accompanying manifold are not closed ones. Condition (6.1) relates 
to the system tied to the accompanying manifold, whereas
condition (6.2) may relate only to the coordinate system that is tied to
a pseudostructure. The second condition in formula (6.2) being the equation
for the pseudostructure points to this fact.
It is only the locally inertial coordinate system that can be such
a coordinate system. From this it follows that
{\it the degenerate transform is realized as the transition from the
accompanying noninertial coordinate system to the locally inertial that}.

To the degenerate transform it must correspond  vanishing of some functional 
expressions. As it was already mentioned in Section 3, 
such functional expressions may be Jacobians, determinants, the Poisson
brackets, residues and others. Vanishing of these
functional expressions is the closure condition for a dual form. 

The conditions of degenerate transform are connected with symmetries 
that can be obtained from the coefficients of evolutionary and dual forms 
and their derivatives. Since the evolutionary relation describes the material 
systems and the coefficients of the form
$\omega^p$ depend on the material system characteristics, it is obvious that
the additional condition (the condition of degenerate transform) has to be due 
to the material system properties. This may be, for example, the availability 
of any degrees of freedom in the system. The translational degrees of freedom,
internal degrees of freedom of the system elements, and so on can be examples
of such degrees of freedom.

The availability of the degrees of freedom in the material system indicates
that it is allowed the degenerate transform, which, in turns, allows
the state of the material systems to be transformed from a nonequilibrium
state to a locally equilibrium state. But, for this to take 
place in reality it is necessary that the additional conditions connected 
with the degrees of freedom of the material system be realized. It is 
selfvariation of the nonequilibrium state of the material system described
by the selfvarying evolutionary relation that could give rise to realization 
of the additional conditions. This can appear only spontaneously because 
it is caused by internal (rather than external) reasons (the degrees of
freedom are the characteristics of the system rather than of external actions).

\section{The origination of the physical structures. 
Formation of physical fields and manifolds. The causality.}
The identical relation (6.3) obtained from the nonidentical evolutionary 
relation 
(under degenerate transform) integrates the state differential and the closed
inexact exterior form. The availability of the state differential indicates
that the material system state becomes a locally equilibrium state (that is, 
the local domain of the system under consideration changes
into the equilibrium state). The availability of the exterior closed
inexact form means that the physical structure is present. This shows that
the transition of material system into the locally equilibrium state is
accompanied by the origination of physical structures. 

Since closed inexact exterior forms corresponding to physical structure
are obtained from the evolutionary relation for the material system, it follows
that physical structures are generated by the material systems. (This is
controlled by the conservation laws.)

In this manner the crated physical structures are connected with the material
system, its elements, its local domains.

In the material system  origination of a physical structure reveals as a new
measurable and observable formation that
spontaneously arises in the material system. 
\{{\it As the examples it can be fluctuations, pulsations, 
waves, vortices, creating massless particles.}\}.  

In the physical 
process this formation is spontaneously extracted from the local
domain of the material system and so it allows the local domain of material
system to get rid of an internal 
force and come into a locally equilibrium state.

The formation created in a local domain of the material system
(at the cost of an unmeasurable quantity that acts in the local domain
as an internal force)
and liberated from that, begins acting onto the neighboring local domain
as a force. This is a potential force, this fact is indicated by the double
meaning of the closed exterior form (on the one hand, a
conservative quantity, and, on other hand, a potential force). 
(This action was produced by the material system in itself,
and therefore this is a potential action rather than an arbitrary one).
The transition of the material system from nonequilibrium into
a locally equilibrium state (which is indicated by the formation of
a closed form) means that the unmeasurable quantity described by the nonzero 
commutator of the nonintegrable differential form $\omega^p$, that acts as an
internal force, transforms into the measurable quantity. It is evident that 
it is just the measurable quantity that acts as a potential force. In other 
words, the internal force transforms into a potential force.

The neighboring domain of the material system works over this action
that appears to be external with respect to that. If in the process the
conditions of conjugacy of the balance conservation laws turn out to be
satisfied again, the neighboring domain
will create a formation by its own, and this formation will be extracted
from this domain. In such a way the formation can move
relative to the material system. (Waves are the example of such motions).

\subsection*{Characteristics of a created formation: intensity, vorticity, 
absolute and relative speeds of propagation of the formation.}
Analysis of the formation and its characteristics allows
a better understanding of the specific features of physical structures.

It is evident that the characteristics of the formation, as well as 
those of the created physical structure, are determined by the evolutionary 
form and its commutator and by the material system characteristics.

The following correspondence between the characteristics of the
formations emerged and characteristics of the evolutionary forms, of the 
evolutionary form commutators and of the material system is established:

1) an intensity of the formation (a potential force)
$\leftrightarrow$ {\it the  value of the first term in the
commutator of a nonintegrable form} at the instant of the formation
is created;

2) vorticity $\leftrightarrow$ {\it the second term in the commutator
that is connected with the metric form commutator};

3) an absolute speed of propagation of the created formation (the
speed in the inertial
frame of reference) $\leftrightarrow$ {\it additional
conditions connected with degrees of freedom of the material system};

4) a speed of the formation propagation relative to the material system
$\leftrightarrow $  {\it additional
conditions connected with degrees of freedom of the material system
and the velocity of the local domain elements}.

\subsection*{Characteristics of physical structures} 
Analysis of formations originated in the material system that correspond to
physical structures allows us to clarify some properties of physical 
structures.

The observed formation and the physical structure are not identical objects.
If the wave be such a formation, the structures formed by the wave front elements 
while its translation are examples of the physical structure.  (In this 
case the wave element is a minipseudostructure. Under this translation 
such element forms a line or some surface, which are examples of the pseudostructure.) 
The observed formations are assigned to material systems, and the physical structures do to 
physical fields. In this case they are mutually connected. As it was already noted, 
the characteristics of physical structure, as well as the characteristics of 
the formation, are determined by the evolutionary 
form and its commutator and by the material system characteristics. 

The originated physical structure  is a realized pseudostructure and
a corresponding inexact closed exterior form. What determines the
characteristics of these objects?

The equation of the pseudostructure is obtained from the condition 
of degenerate transform and is dictated by a degree of freedom 
of material system. The pseudostructure is connected with the absolute speed 
of propagation of 
the created formation (see point 3 in the preceding section). 

A closed exterior form corresponding to a physical structure is a conservative
quantity that the state differential corresponds to. The differentials of 
entropy, action, potential and others are the examples of such differentials. 
These conservative quantities describe certain charges.

A physical structure possesses two more characteristics.
This is connected with the fact that {\it inexact} closed forms correspond
to these structures. Under transition from one structure to another
the conservative quantity corresponding to the closed exterior form discretely
changes, and the pseudostructure also changes discretely.

Discrete changes of the conservative quantity and pseudostructure are 
determined by the value of the evolutionary form commutator, which the 
commutator has at
the instant when the physical structure originates. The first term of the
evolutionary form commutator obtained from the derivatives of the evolutionary 
form coefficients controls the discrete change of the conservative
quantity. As a corresponding characteristics of the created formation it serves 
the intensity of  created formation (see point 1).  
The second term of the evolutionary form commutator obtained from the 
derivatives of the metric form coefficients of the initial manifold 
controls the pseudostructure change. Spin is the example of the second 
characteristics. As a corresponding characteristics of the created formation it serves 
the vorticity of  created formation (see point 2). Spin is a characteristics 
that determines a character of the manifold deformation before origination of
the quantum. (The spin value depends on the form degree.)

A discrete change of the conservative quantity and that of the pseudostructure
produce the quantum that is obtained while going from one structure to another.
The evolutionary form commutator formed at the instant of the structure
origination determine characteristics of this quantum. 

Breaks of the derivatives of the potential along the direction normal
to the potential surface, breaks of the derivative in transition
throughout the characteristic surfaces and in transition throughout the
wave front, and others are the examples of such discrete changes.

The duality  of the closed inexact form as a conservative
quantity and as a potential force shows that
the potential forces are the action of formations corresponding to
the physical structures onto the material system elements.

Here the following should be pointed out. The physical structures are 
generated by local domains of the material system. They are the elementary 
physical structures. By combining with one another they can form the
large-scale structures and physical fields.

\subsection*{Classification of physical structures. Formation of physical fields.
(Parameters of the closed and dual forms)}
Since the physical structures are generated by numerous local domains of the
material system and at numerous instants of realizing various degrees
of freedom of the material system, it is evident that they can generate fields.
In this manner physical fields are formatted.
To obtain the physical structures that form a given physical field one has to
examine the material system  corresponding to this field and the appropriate 
evolutionary relation. In particular, to obtain the thermodynamic
structures (fluctuations, phase transitions, etc) one has to analyze the
evolutionary relation for the thermodynamic systems, 
to obtain the gas dynamic ones (waves, jumps, vortices, pulsations) 
one has to employ the evolutionary relation for gas dynamic 
systems, for the electromagnetic field one must employ
a relation obtained from equations for charged particles.

Closed forms that correspond to physical structures are generated by
the evolutionary relation having the parameter $p$ that defines a number of
interacting balance conservation laws, that ranges from 0 to 3. Therefore, the physical structures
can be classified by the parameter $p$. The other parameter is a degree
of  closed forms generated by the evolutionary relation.
As it was shown above, the evolutionary relation of
degree $p$ can generate the closed forms of degree $0\leq k \leq p$. Therefore,
physical structures can be classified by the parameter $k$ as well.

In addition, closed exterior forms of the same degree realized in spaces of
different dimensions prove to be distinguishable because the dimension of the
pseudostructures, on which the closed forms are defined, depends on the space
dimension. As a result, the space dimension also specifies the physical
structures. This parameter determines the properties of the physical structures
rather than their type.

What is implied by the concept ``space"?

While deriving the evolutionary relation two frames of reference were
used and, correspondingly, two spatial objects . The first frame of reference
is the inertial one that is connected with the space where the material 
system is situated and is not directly connected with the material system. 
This is an inertial space, it is a metric space. (As it will be shown below,
this space is also formed by the material system itself.) The second frame of
reference is the proper one, it is connected with the accompanying manifold, 
which is not a metric manifold.

If the dimension of the initial inertial space is $n$, then while  
generating closed forms of sequential degrees 
$k=p$, $k=p-1$, \dots, $k=0$ the pseudostructures of dimensions
$(n+1-k)$: 1, \dots, $n+1$ are obtained. As a result of transition
to the exact closed form of zero degree the metric structure of the
dimension $n+1$ is obtained. Under the influence of an external action
(and in the presence of degrees of freedom) the material system connected with 
the accompanying manifold can transfer the initial inertial space into the 
space of the dimension $n+1$. $\{$It is known that the skew-symmetric tensors 
of the rank $k$ correspond to the closed exterior differential forms, and the 
pseudotensors of the rank $(N-k)$, where $N$ is the space dimension, correspond 
to the relevant dual forms. The pseudostructures correspond to such tensors, 
but on the space formed with the dimension $n+1$. That is, $N=n+1$\}. 

Hence, 
from the analysis of the evolutionary relation one can see that the type 
and the properties of the physical structures (and, accordingly, of physical
fields) for a given material system depend on a number of interacting balance
conservation laws $p$, on the degree of realized closed forms $k$, and on the
space dimension. 

The connection between the field theory equations and closed exterior forms 
presented in Section 1 shows that it is possible to classify physical fields
according to the degree of exterior differential form. But within the framework 
of only exterior differential forms one cannot understand how this classification 
is explained. This can be elucidated only by application of evolutionary 
differential forms. 

By introducing a classification with respect to $p$, $k$,
and a space dimension we can understand the internal connection of various
physical fields and interactions. 
So one can see a correspondence between the degree $k$ of the
closed forms realized and the type of interactions. Thus, $k=0$ corresponds to
the strong interaction, $k=1$ corresponds to the weak interaction,
$k=2$ corresponds to the electromagnetic interaction, and $k=3$ corresponds
to the gravitational interaction.

\subsection*{Causality}
The mathematical
apparatus of evolutionary differential forms describing the
balance conservation laws in the material systems explains the causality 
of the evolutionary processes in material systems, which lead to emergence of 
physical structures. 

The evolutionary process in the material system and the emergence of physical
structures can take place, if

1) the material system is subjected to an external action ({\it the 
evolutionary form commutator in the evolutionary relation obtained from the 
balance conservation laws is nonzero}),

2) a material system possesses the degrees of freedom ({\it there
are the conditions of degenerate transform, under which from the
nonidentical evolutionary relation an identical relation is obtained}),

3) the degrees of freedom of the material system have to be realized,
that is possible only under
selfvariations of the nonequilibrium state of material system ({\it
the conditions of the degenerate transform
have to be satisfied, this is possible under selfvariation of the
nonidentical evolutionary relation}).

If these conditions (causes) are fulfilled, in the material system
physical structures arise ({\it this is indicated by the presence of closed
inexact form obtained from the identical relation}).

It is these structures that form physical fields.

Note that to each physical field it is assigned its own material system. 
The question of what physical system corresponds 
to a particular physical field is still an open question. Examples of such
material systems are the thermodynamic, gas dynamical, the system of charged 
particles, cosmological systems, and so on. Maybe, for elementary particles
the physical vacuum is such a system.

The evolutionary process may lead to creation of elements of its own
material system ({\it while obtaining the exact form of zero degree}).
(In this case the numbers are the analog of dual forms).

The emergence of physical structures in the evolutionary process proceeds
spontaneously and is manifested as an emergence of certain observable
formations. In this manner the causality of emerging various
observable formations in material media is explained. Such formations and their 
manifestations are fluctuations, turbulent pulsations, waves, vortices, 
creating massless particles and others.

Here the role of the conservation laws in evolutionary processes should be 
emphasized once again. The noncommutativity of the balance conservation laws
leads to the nonequilibrium state of the material system (subjected to external 
actions). The interaction of the noncommutative balance
conservation laws controls the process of selfvarying
the nonequilibrium state of the material system that leads to the
realization of the degrees of freedom of the material system. 
This allows the material system to transform into the locally equilibrium state,
which is accompanied by emergence of physical structures. The exact
conservation laws correspond to the physical structures. Here one can see a 
connection between the balance and exact
conservation laws. The physical structures that correspond to the exact
conservation laws are produced by material system in the evolutionary
processes based on the interaction of the noncommutative balance conservation
laws [31].

{\it The noncommutativity of the balance conservation laws
and their controlling role in the evolutionary processes, that are
accompanied by emerging  physical structures, practically
have not been taken into account in the explicit form  anywhere}. The 
mathematical apparatus of evolutionary differential forms enables one to take
into account and describe these points.

\bigskip
The existing field theories that are invariant ones are based on some 
postulates. The investigation performed allows us to make the following 
conclusion. The postulates, which lie at the basis of the existing field 
theories, correspond to the statement about the conjugacy of the
balance conservation laws for material systems that generate physical
structures forming physical fields.

It can be shown that any material system and any physical field is subjected to 
the above described regularities, which were obtained with thw help of the 
mathematical apparatus of the exterior and evolutionary differential forms.  
Let us consider some material systems and physical fields.

\bigskip
1. {\it The thermodynamic systems}

The thermodynamics is based on the first and second principles of thermodynamics
that were introduced as postulates [15].
The first principle of thermodynamics, which can be written in the form
$$dE\,+\,dw\,=\,\delta Q\eqno(7.1)$$
follows from the balance conservation laws for energy and linear momentum 
(but not only from the conservation law for energy). This is analogous to the evolutionary relation for 
the thermodynamic system. Since $\delta Q$ is not a differential,
relation (7.1) which corresponds to the first principle of thermodynamics, as 
well as the evolutionary relation, appears to be a nonidentical relation. 
This points to a noncommutativity of the balance conservation
laws (for energy and linear momentum) and to a nonequilibrium state of the
thermodynamic system.

If condition of the integrability be satisfied, from the nonidentical 
evolutionary relation, which corresponds to the first principle of 
thermodynamics, it follows an identical relation. It is an identical relation
that corresponds to the second principle of thermodynamics.

If $dw\,=\,p\,dV$,  there is the integrating factor 
$\theta$ (a quantity which depends only on the characteristics of the system),
where $1/\theta\,=\,pV/R$ is called the temperature $T$ [15]. 
In this case the form $(dE\,+\,p\,dV)/T$ turns out to be a differential 
(interior) of some quantity that referred to as entropy $S$:
$$(dE\,+\,p\,dV)/T\,=\,dS \eqno(7.2)$$

If the integrating factor $\theta=1/T$ has been
realized, that is, relation (7.2) proves to be satisfied, from relation (7.1), 
which corresponds to the first principle of thermodynamics,
it follows
$$dS\,=\,\delta Q/T \eqno(7.3)$$
This is just the second principle of thermodynamics for reversible processes.
It takes place when the heat influx is the only action onto the system.

If in addition to the heat influx the system experiences a certain mechanical
action 
(for example, an influence of boundaries), then we obtain
$$dS\, >\,\delta Q/T \eqno (7.4)$$
that corresponds to the second principle of thermodynamics for irreversible
processes.

In the case examined above a differential of entropy (rather than entropy 
itself) becomes a closed form. $\{$In this case entropy  manifests itself 
as the thermodynamic potential, namely, the function of state. To the
pseudostructure there corresponds
the state equation that determines the temperature dependence on the
thermodynamic variables$\}$.

\bigskip
2. {\it The gas dynamical systems}

We take the simplest gas dynamical system, namely, a flow
of ideal (inviscous, heat nonconductive) gas [23].

Assume that gas (the element of gas dynamic system) is a thermodynamic system in 
the state of local equilibrium (whenever the gas dynamic system itself may be 
in nonequilibrium state), that is, it is satisfied the relation [15]
$$Tds\,=\,de\,+\,pdV \eqno(7.5)$$
where $T$, $p$ and $V$ are the temperature, the pressure and the gas 
volume, $s$, $e$ are entropy and internal energy per unit volume.

Let us introduce two frames of reference: an inertial one that is not connected 
with the material system and an accompanying frame of reference that is connected 
with the manifold formed by the trajectories of the material system elements. 

The equation of the balance conservation law of energy for ideal gas can
be written as [23]
$${{Dh}\over {Dt}}- {1\over {\rho }}{{Dp}\over {Dt}}\,=\,0 \eqno(7.6)$$
where $D/Dt$ is the total derivative with respect to time (if to designate
the spatial coordinates by $x_i$ and the velocity components by $u_i$,
$D/Dt\,=\,\partial /\partial t+u_i\partial /\partial x_i$). Here  $\rho=1/V $
and $h$ are respectively the mass and the entalpy densities of the gas.

Expressing entalpy in terms of internal energy $e$ with the help of formula
$h\,=\,e\,+\,p/\rho $ and using relation (7.5) the balance conservation law
equation (7.6) can be put to the form
$${{Ds}\over {Dt}}\,=\,0 \eqno(7.7)$$

And respectively, the equation of the balance conservation law for linear
momentum can be presented as [23,27]
$$\hbox {grad} \,s\,=\,(\hbox {grad} \,h_0\,+\,{\bf U}\times \hbox {rot} {\bf U}\,-{\bf F}\,+\,
\partial {\bf U}/\partial t)/T \eqno(7.8)$$
where ${\bf U}$ is the velocity of the gas particle, 
$h_0=({\bf U \cdot U})/2+h$, ${\bf F}$ is the mass force. The operator $grad$ 
in this equation is defined only in the plane normal to the trajectory.

Since the total derivative with respect to time is that along the trajectory,
in the accompanying frame of reference equations (7.7) and (7.8)
take the form:
$${{\partial s}\over {\partial \xi ^1}}\,=\,0 \eqno (7.9)$$
$${{\partial s}\over {\partial \xi ^{\nu}}}\,=\,A_{\nu },\quad \nu=2, ... \eqno(7.10)$$
where $\xi ^1$ is the coordinate along the trajectory,
$\partial s/\partial \xi ^{\nu }$
is the left-hand side of equation (7.8), and $A_{\nu }$ is obtained from the
right-hand side of relation (7.8). 

Equations (7.9) and (7.10) can be convoluted into the equation
$$ds\,=\,A_{\mu} d\xi ^{\mu}\eqno(7.11)$$
where $\,A_{\mu} d\xi ^{\mu}=\omega\,$ is the first degree differential form
(here $A_1=0$,$\mu =1,\,\nu $).

Relation (7.11) is an evolutionary relation for gas dynamic system
(in the case of local thermodynamic equilibrium). Here $\psi\,=\,s$.
$\{$It worth notice that in the evolutionary relation for thermodynamic
system the dependence of entropy on thermodynamic variables is investigated
(see relation (7.5)), whereas in the evolutionary relation for gas dynamic
system the entropy dependence on the space-time variables is considered$\}$.

Relation (7.11) appears to be nonidentical. To make it sure that this is true 
one must inspect a commutator of the form $\omega $. 

Nonidentity of the evolutionary relation points to the nonequilibrium and a 
development of the gas dynamic instability.   
Since the nonequilibrium is produced by internal forces that are described
by the commutator of the form $\omega $, it becomes evident that a cause
of the gas dynamic instability is something that contributes into the
commutator of the form $\omega $. 

One can see (see (7.8)) that the development of instability is caused by
not a simply connectedness of the flow domain,  nonpotential  external
(for each local domain of the gas dynamic system) forces, a nonstationarity
of the flow. 

\{In the case when gas is
nonideal equation (7.9) can be written in the form
$${{\partial s}\over {\partial \xi ^1}} \,=\,A_1 \eqno$$
where $A_1$ is an expression that depends on the energetic actions (transport 
phenomena: viscous, heat-conductive). 
In the case of reacting gas extra terms connected with the chemical
nonequilibrium are added. These factors contributes into the
commutator of the form $\omega $.\} 

All these factors lead to emergence of internal forces, 
that is, to nonequilibrium and to development
of various types of instability. 

And yet for every
type of instability one can find an appropriate term giving contribution
into the evolutionary form commutator, which is responsible for this type
of instability.
Thus, there is an unambiguous connection between the type of instability
and the terms that contribute into  the evolutionary form commutator in the
evolutionary relation. \{In the general case one has to consider the
evolutionary relations that correspond to the balance conservation laws 
for angular momentum and mass as well\}.

As it was shown above, under realization of additional degrees of freedom
it can take place the transition from the nonequilibrium state to the locally
equilibrium one, and this process is accompanied by emergence of physical
structures. 
The gas dynamic formations that correspond to these physical structures are
shocks, shock waves, turbulent pulsations and so on. Additional degrees of 
freedom are realized as the condition of the degenerate transform, namely, 
vanishing of determinants, Jacobians of transforms, etc. These conditions 
specify the integral surfaces (pseudostructures):
the characteristics (the determinant of coefficients at the normal derivatives
vanishes), the singular points (Jacobian is equal to zero), the envelopes
of characteristics of the Euler equations and so on. Under passing
throughout the integral surfaces
the gas dynamic functions or their derivatives suffer breaks. 

\bigskip
3. {\it Electromagnetic field}
 
The system of charged particles is a material medium, which 
generates  electromagnetic field. 

If to use the Lorentz force ${\bf F\,= \,\rho (E + [U\times H]}/c)$,
the local variation of energy and linear momentum of the charged
matter (material system) can be written respectively as [24]: $\rho ({\bf U\cdot E})$,
$\rho ({\bf E+[U\times H]}/c)$. Here $\rho$ is the charge
density, ${\bf U}$ is the velocity of the charged matter. These
variations of energy and linear momentum are caused by energetic and
force actions and are equal to values of these actions. If to denote
these actions by $Q^e$, ${\bf Q}^i$, the balance conservation laws
can be written as follows:
$$\rho \,({\bf U\cdot E})\,=\,Q^e\eqno(7.12)$$
$$\rho \,({\bf E\,+\,[U\times H]}/c)\,=\, {\bf Q}^i \eqno(7.13)$$

After eliminating  the characteristics of the material system (the charged
matter) $\rho$ and ${\bf U}$ by application of the Maxwell-Lorentz equations
[24], the left-hand sides of equations (7.12), (7.13) can be expressed only
in terms of the strengths of electromagnetic field, and then one can write
equations (7.12), (7.13) as
$$c\,\hbox{div} {\bf S}\,=\,-{{\partial}\over {\partial t}}\,I\,+\,Q^e\eqno(7.14)$$
$${1\over c}\,{{\partial }\over {\partial t}}\,{\bf S}\,=
\,{\bf G}\,+\,{\bf Q^i}\eqno(7.15)$$
where ${\bf S=[E\times H]}$ is the Pointing vector, $I=(E^2+H^2)/c$,
${\bf G}={\bf E}\,\hbox {div}{\bf E}+\hbox{grad}({\bf E\cdot E})-
({\bf E}\cdot \hbox {grad}){\bf E}+\hbox {grad}({\bf H\cdot H})-({\bf H}\cdot\hbox{grad}){\bf H}$.

Equation (7.14) is widely used while describing electromagnetic
field and calculating  energy and the Pointing vector. But equation (7.15)
does not commonly be taken into account. Actually, the Pointing vector
${\bf S}$ must obey two equations that can be convoluted into the
{\it relation}
$$d\bf S=\,\omega ^2\eqno(7.16)$$
Here $d\bf S$ is the state differential being 2-form and the coefficients
of the form $\omega ^2$ (the second degree form) are the right-hand sides
of equations (7.14), (7.15).
It is just the evolutionary relation for the system of charged particles that
generate electromagnetic field.

By analyzing the coefficients of the form $\omega ^2$ (obtained from equations
(7.14), (7.15), one can assure oneself that the form commutator is nonzero.
This means that from relation (7.16) the Pointing vector cannot be found.
This points to the fact that there is no such a measurable quantity
(a potential).

Under what conditions
can the Pointing vector be formed as a measurable quantity?

Let us choose the local coordinates $l_k$ in such a way that one direction
$l_1$ coincides with the direction of the vector ${\bf S}$. Because this
chosen direction coincides with the direction of the vector
${\bf S=[E\times H]}$ and hence is normal to the vectors
${\bf E}$ and ${\bf H}$,
one obtains that $\hbox{div} {\bf S}\,=\,\partial s/\partial l_1$,
where $S$ is a module of ${\bf S}$. In addition,
the projection of the vector ${\bf G}$ on the chosen direction turns out to be
equal to $-\partial I/\partial l_1$.
As a result, after separating from vector equation (7.15) its projection
on the chosen direction equations (7.14), (7.15) can be written as
$${{\partial S}\over {\partial l_1}}\,=\,-{1\over c}{{\partial I}\over {\partial t}}\,+\,
{1\over c}Q^e \eqno(7.17)$$
$${{\partial S}\over {\partial t}}\,=\,-c\,{{\partial I}\over {\partial l_1}}\,+\,c{\bf Q}'^i\eqno(7.18)$$
$$0\,=\,-{\bf G}''\,-\,c{\bf Q}''^i$$
Here the prime relates to the direction $l_1$, double primes relate to
the other directions. Under the condition $d l_1/d t\,=\,c$ from
equations (7.17), (7.18) it is possible to obtain the relation in differential
forms
$${{\partial S}\over {\partial l_1}}\,dl_1\,+\,{{\partial S}\over {\partial t}}\,dt\,=\,
-\left( {{\partial I}\over {\partial l_1}}\,dl_1\,+\,{{\partial I}\over {\partial t}}\,dt\right )\,+\,
(Q^e\,dt\,+\,{\bf Q}'^i\,dl_1)\eqno(7.19)$$
Because the expression within the second braces in the right-hand side is
not a differential (the energetic and force
actions have different nature and cannot be conjugated), one can obtain
a closed form only if this term vanishes:
$$(Q^e\,dt\,+\,{\bf Q}'^i\,dl_1)\,=\,0\eqno(7.20)$$
that is possible only discretely (rather than identically).

In this case $dS\,=\,0$, $dI\,=\,0$ and the modulus of the Pointing vector $S$ 
proves to be a closed form, i.e. a measurable quantity. The integrating 
direction (the pseudostructure) will be
$$-\,{{\partial S/\partial t}\over {\partial S/\partial l_1}}\,=\,{{dl_1}\over {dt}}\,=\,c\eqno(7.21)$$
The quantity $I$ is the second dual invariant.

Thus, the constant $c$ involving into the Maxwell equations is defined
as the integrating direction.

\bigskip
4. {\it Gravitational field} 

Material system (medium), which generates 
gravitational field, is cosmological system.  
What can be said about the pseudo-Riemann manifold and Riemann space? 
The distinctive property of the Riemann manifold is an availability of 
the curvature. This means that the metric form commutator of the third 
degree is nonzero. Hence, it does not equal zero the evolutionary form 
commutator of the third degree $p=3$, which involves into itself the metric 
form commutator. That is, the evolutionary form that enters into the 
evolutionary relation is unclosed, and the relation is nonidentical. 

When realizing pseugostructures of the dimensions $1$, $2$, $3$, and $4$ and 
obtaining the closed inexact forms of the degrees $k=3$, $k=2$, $k=1$, $k=0$, 
the pseudo-Riemann space is formed, and the transition to the exact 
form of zero degree corresponds to the transition to the Riemann space. 

It is well known that while obtaining the Einstein equations it was 
suggested that there are fulfilled the conditions [28]: the Bianchi identity 
is satisfied, the coefficients of connectedness are symmetric, the 
condition that the coefficients of connectedness are the Christoffel 
symbols, and an existence of the transformation, under which the 
coefficients of connectedness vanish. These conditions are the conditions 
of realization of the degenerate transformations for nonidentical relations 
obtained from the evolutionary relation of the degree $p=3$ and after going to the exact relations. In this case to the Einstein equation there corresponts the identical equation of the first degree.

\bigskip
In conclution we can mention the following.
 
Physical fields are described by invariant field theory that is based on exact 
conservation laws. The properties of closed exterior differential forms lie at 
the basis of mathematical apparatus of the invariant theory. The 
mechanism of {\it forming } physical fields can be described only by 
evolutionary theory. The evolutionary theory that is based on 
the balance conservation laws for material systems is just such a theory. 
It is evident that as the common 
field theory it must serve a theory that involves the basic mathematical 
foundations of the evolutionary and invariant field theories.

1. Cartan E., Lecons sur les Invariants Integraux. -Paris, Hermann, 1922.  

2. Cartan E., Les Systemes Differentials Exterieus ef Leurs Application 
Geometriques. -Paris, Hermann, 1945.  

3. Finikov S.~P.,  Method of the Exterior Differential Forms by Cartan in 
the Differential Geometry. Moscow-Leningrad, 1948 (in Russian).

4. Sternberg S., Lectures on Differential Geometry. -Englewood Cliffd , N.J.:
Prentiice-Hall,1964

5. Encyclopedia of Mathematics. -Moscow, Sov.~Encyc., 1979 (in Russian).

6. Bott R., Tu L.~W., Differential Forms in Algebraic Topology. 
Springer, NY, 1982.

7. Petrova L.~I., Exterior differential forms in the field theory. 
//Abstracts of International Conference dedicated to 
the 90$^{th}$ Anniversary of L.~S.~Pontryagin, 
Algebra, Geometry, and Topology. Moscow, 1998, 123-125 (in Russian).

8. Petrova L.~I., The role of the conservation laws in evolutionary processes. 
Spacetime and Substance, 1, 2001, 1-23

9. Schutz B.~F., Geometrical Methods of Mathematical Physics. Cambrige 
University Press, Cambrige, 1982.

10. Weinberg S., Gravitation and Cosmology. Principles and applications of 
the general theory of relativity. Wiley \& Sons, Inc., N-Y, 1972.

11. Novikov S.~P., Fomenko A.~P., Elements of the differential geometry and 
topology. -Moscow, Nauka, 1987 (in Russian). 

12. Wheeler J.~A., Neutrino, Gravitation and Geometry. Bologna, 1960.

13. Konopleva N.~P. and Popov V.~N., The gauge fields. Moscow, Atomizdat, 1980 
(in Russian).

14. Smirnov V.~I., A course of higher mathematics. -Moscow, 
Tech.~Theor.~Lit. 1957, V.~4 (in Russian).

15. Haywood R.~W., Equilibrium Thermodynamics. Wiley Inc. 1980.

16. Fock V.~A., Theory of space, time, and gravitation. -Moscow, 
Tech.~Theor.~Lit., 1955 (in Russian).

17. Efimov N.~V. Exterior Differntial Forms in the Euclidian Space. 
Moscow State Univ., 1971.

18. Tonnelat M.-A., Les principles de la theorie electromagnetique 
et la relativite. Masson, Paris, 1959.

19. Vladimirov V.~S., Equations of the mathematical physics. -Moscow, 
Nauka, 1988 (in Russian).

20. Dirac P.~A.~M., The Principles of Quantum Mechanics. Clarendon Press, 
Oxford, UK, 1958.

21. Encyclopedic dictionary of the physical sciences. -Moscow, Sov.~Encyc., 
1984 (in Russian).

22. Dafermos C.~M. In "Nonlinear waves". Cornell University Press, 
Ithaca-London, 1974.

23. Clark J.~F., Machesney ~M., The Dynamics of Real Gases. Butterworths, 
London, 1964.

24. Tolman R.~C., Relativity, Thermodynamics, and Cosmology. Clarendon Press, 
Oxford,  UK, 1969.

25. Petrova L.~I., Origination of physical structures. //Izvestia RAN, Fizika, N 1, 2003.  

26. Prigogine I., Introduction to Thermodynamics of Irreversible 
Processes. --C.Thomas, Springfild, 1955.

27. Liepman H.~W., Roshko ~A., Elements of Gas Dynamics. Jonn Wiley, 
New York, 1957

28. Einstein A. The Meaning of Relativity. Princeton, 1953.

\end{document}